\documentclass[11pt]{amsart}
\usepackage{latexsym,amssymb}
\usepackage{graphics}
\usepackage{harvard}
\usepackage{enumerate}
\usepackage{color}
\usepackage{url}
\usepackage{graphicx,subfig}
\usepackage{setspace}
\usepackage{soul}
\usepackage{multirow}
\newcommand\N{\mathbb{N}}

\newcommand\R{\mathbb{R}}

\newcommand\eps{\varepsilon}
\newcommand\E{\mathbb{E}}     

\newcommand\Fc{\mathcal{F}}
\newcommand\Gc{\mathcal{G}}

\newcommand\Dc{\mathcal{D}}

\newcommand\Nc{\mathcal{N}}

\newcommand\Gb{\mathbb{G}}

\newcommand\Fb{\mathbb{F}}

\DeclareMathOperator{\Var}{Var}

\DeclareMathOperator{\Cov}{Cov}

\newcommand\weak{\ \rightsquigarrow\ }

\newcommand\deq{ \stackrel{\Dc}{=} }

\newtheoremstyle{normal}
{2ex}               
{3ex}               
{}                  
{}                  
{\bfseries} 
{}                  
{2pt}   
{\thmname{#1}\thmnumber{ #2.} \thmnote{(#3)}}

\newtheoremstyle{italic}
{2ex}
{3ex}
{\itshape}
{}
{\bfseries} 
{}
{2pt}
{\thmname{#1}\thmnumber{ #2.} \thmnote{(#3)}}

\theoremstyle{italic}
\newtheorem{definition}{Definition}[section]

\newtheorem{theorem}[definition]{Theorem}

\newtheorem{prop}[definition]{Proposition}

\theoremstyle{normal}
\newtheorem{remark}[definition]{Remark}
\newtheorem{example}[definition]{Example}
\usepackage[margin=1.1in]{geometry} 

\def\RR{\mathbb{R}}

\def\EE{\mathbb E}

\def\SS{\mathcal S}
\def\GG{\mathcal G}

\def\NN{\mathcal N}

\def\Cramer{Cram\'er }

\begin{document}
\bibliographystyle{econometrica}
\title{Testing for Homogeneity in Mixture Models}

\author{Jiaying Gu}
\author{Roger Koenker}
\author{Stanislav Volgushev}
\thanks{Version:  \today .   This research was partially supported by NSF grant SES-11-53548 and Project C1 of the SFB 823 of the German Research Foundation. Part of this research was conducted while the first author was visiting the Mathematics department at Ruhr University Bochum and the third author was a visiting scholar at UIUC. They are very grateful to the UIUC Statistics and Economics departments and the Bochum Mathematics department for their hospitality. The third author also gratefully acknowledges Financial support from the DFG (grant VO1799/1-1). The authors would also like to express their appreciation to
the Co-Editor and the referees for comments that led to improvements in the paper.
}

\begin{abstract}
  Statistical models of unobserved heterogeneity are typically formalized as
  mixtures of simple parametric models and interest naturally focuses on 
  testing for homogeneity versus general mixture alternatives.  Many tests of
  this type can be interpreted as $C(\alpha)$ tests, as in \citeasnoun{Neyman59},
  and shown to be locally, asymptotically optimal.  These $C(\alpha)$ tests 
  will be contrasted with a new approach to likelihood ratio testing for general
  mixture models.  The latter tests are based on estimation of general
  nonparametric mixing distribution with the \citeasnoun{KW} maximum likelihood
  estimator.  Recent developments in convex optimization have dramatically
  improved upon earlier EM methods for computation of these estimators, and
  recent results on the large sample behavior of likelihood ratios involving
  such estimators yield a tractable form of asymptotic inference. Improvement in 
  computation efficiency also facilitates the use of a bootstrap methods to 
  determine critical values that are shown to work better than the asymptotic 
  critical values in finite samples. Consistency of the bootstrap procedure 
  is also formally established. We compare performance of the two approaches 
  identifying circumstances in which each is preferred.  
\end{abstract}
\maketitle
\pagestyle{myheadings}
\markboth{\sc Inference for Mixture Models}{\sc Gu, Koenker and Volgushev}
\section{Introduction}

Given a simple parametric density model, $p(x|\mu)$, for iid observations, 
$X_1, \cdots , X_n$, there is a natural temptation to complicate the model by
allowing the parameter, $\mu$, to vary with the observation index.  In the absence of other,
e.g. observable covariate, information that would distinguish the observations from one another
it may be justifiable to view the $\mu$'s as drawn at random.  Inference for such mixture
models is complicated by the enormous class of potential alternatives.
Two dominant approaches to testing for homogeneity in such models exist:  
Neyman's $C(\alpha)$ tests and likelihood ratio tests.
$C(\alpha)$ tests are particularly attractive for testing homogeneity  since like
their kindred score tests they do not require estimation of the model under the
alternative of heterogeneity of the parameter $\mu$. As described in \citeasnoun{Gu}, 
$C(\alpha)$ tests have a somewhat irregular, 
but still relatively simple asymptotic theory, and are generally easy to compute.  
Likelihood ratio tests, in contrast, are known to have a considerably more complicated limiting behavior, 
and are generally regarded as much more difficult to compute.  Our primary objective here
is to try to rehabilitate the reputation of the LRT for testing homogeneity in mixture models
by demonstrating that it is both computationally tractable and  -- at least under some conditions --
that it has attractive power and size control properties when compared to other tests.

We will argue that recent developments in convex optimization have dramatically reduced
the computational burden of the LRT approach for general, nonparametric alternatives.
Following \citeasnoun{Laird}, prior efforts to compute the Kiefer-Wolfowitz MLE for
general nonparametric mixture models have relied upon some variant of the EM algorithm.
However, \citeasnoun{KM} have recently shown that interior point methods for general
convex optimization provide a much more efficient, and more accurate computational approach.
A second impediment to the use of LRT  methods for general mixture problems has been the
lack of a tractable limiting distribution theory.  Extending recent work of 
\citeasnoun{gassiat2002}, \citeasnoun{liushao2003} and \citeasnoun{Azgame2009} 
we propose an easily simulated method of computing limiting
critical values for the LRT statistic for testing homogeneity for general nonparametric 
mixture models. However, we find in simulations that these limiting critical values do 
not serve as a good approximation in moderate samples. Instead we propose a parametric 
bootstrap  method to determine critical values, and formally prove its consistency. 
Size and power performance of the bootstrap method is investigated through simulations.

There is a large and rapidly growing literature on inference for {\it finite} mixture models using 
penalized likelihood ratio methods, which can be considered an intermediate approach between 
$C(\alpha)$ tests  and our general LRT approach based on the Kiefer-Wolfowitz MLE.  Ironically, 
once one restricts mixtures to discrete distributions with a finite number of 
support points, convexity of the log likelihood is lost, making LRT methods considerably
more challenging from a computational point of view. Moreover, finite mixture models fail to satisfy
certain regularity conditions that are typically required for parametric likelihood ratio tests, 
making their asymptotic theory challenging, see for example \citeasnoun{ChoWhite} and \citeasnoun{CPT2014}. 
Motivated by these challenges, \citeasnoun{CCK.01} have proposed penalizing the log likelihood  with a
log barrier penalty on the mixing weights.  The penalty removes the singularity in the log likelihood
that arises when mixing weights tend to zero, and leads to a relatively simple mixture of $\chi^2$ limiting
theory for the restricted LRT statistic.  More recently, \citeasnoun{ChenLi}, 
\citeasnoun{CLM} and \citeasnoun{LiChen} have extended this approach and developed an attractive
inference apparatus for restricted mixture models based on these penalized likelihood ratio methods. 
\citeasnoun{KS2014} further extend the EM test methods to normal mixture regression models.
We will incorporate these EM tests into our performance comparisons in the simulation section of the paper.

The next section provides a detailed discussion of our general approach to likelihood ratio testing
based on the Kiefer-Wolfowitz nonparametric MLE.  The following two sections briefly describe
the $C(\alpha)$ and EM testing approaches.  Simulation evidence on the performance of the various
methods and an empirical example is reported in Section~\ref{sec:sim} and~\ref{sec:eg}.

\section{Likelihood Ratio Tests for Homogeneity in Mixture Models} \label{sec:LRT}
A prerequisite for any likelihood ratio test for general mixture models must be a reliable 
maximum likelihood estimator for these models under the alternative of parameter heterogeneity.
\citeasnoun{Lindsay95} offers a comprehensive overview of the vast literature on mixture
models,  and traces the idea of maximum likelihood estimation of a {\it nonparametric}
mixing measure $\eta$, given random samples from the mixture density,
\begin{equation}
g(x) = \int p(x| \mu ) d\eta(\mu),
\label{Mdens}
\end{equation}
to an {\it Annals} abstract of \citeasnoun{Robbins50}.  Somewhat later \citeasnoun{KW} 
provided a detailed analysis of such a nonparametric MLE and
established its consistency.  Yet only with \citeasnoun{Laird} did a viable computational strategy
emerge for a discretized version.  The EM method proposed by Laird has been employed extensively
in subsequent work, notably by  \citeasnoun{HeckmanSinger} and \citeasnoun{JZ}, even though it has
been widely criticized for its slow convergence.   Recently, \citeasnoun{KM} have
shown that the discretized version Kiefer-Wolfowitz estimator can be formulated as a convex optimization
problem and accurately solved very efficiently by interior point methods.  Recent work by 
\citeasnoun{gassiat2002} and \citeasnoun{Azgame2009} has also clarified the limiting behavior of
the LRT for general classes of alternatives, and taken together these developments
offer a fresh opportunity to explore the viability of the LRT for inference on mixtures.

It seems ironic that many of the difficulties inherent in maximum likelihood estimation
of finite parameter mixture models vanish when we consider nonparametric mixtures.
The notorious multimodality of parametric likelihood surfaces is replaced by a much 
simpler, strictly convex optimization problem possessing a unique solution.  
It is of obvious concern that consideration of such a wide class of alternatives may 
depress the power of associated tests; we will see that while there is some loss of power
when compared to more restricted parametric LRTs, the loss is typically modest,
a small price to pay for power gained against a broader class of alternatives.  We will also see
that by comparison with $C(\alpha)$ tests that are also designed to detect general
alternatives the LRT can be competitive.

\subsection{Maximum Likelihood Estimation of General Mixtures} \label{sec:KWE}
Suppose that we have iid observations, $X_1, \cdots , X_n$  from the mixture density
\eqref{Mdens}, the Kiefer-Wolfowitz MLE requires us to solve,
\[
\min_{\eta \in \bar\GG} \Big\{ - \sum_{i=1}^n \log g(x_i) \Big| g(x_i) = \int p(x_i|\mu) d\eta(\mu) \Big\},
\]
where $\bar\GG$ is the (convex) set of all mixing distributions. The problem is one of
minimizing the sum of strictly convex functions subject to linear equality and inequality constraints.
The dual to this (primal) convex program proves to be somewhat more tractable from a
computational viewpoint, and takes the form,
\[
\max_{\nu \in \RR^n} \Big\{ \sum_{i=1}^n \log \nu_i \; \Big| \; 
\sum_{i=1}^n \nu_i p(x_i | \mu ) \leq n, \quad \mbox{for all} \; \mu \Big\}
\]
See \citeasnoun{LindsayI} and \citeasnoun{KM} for further details.  This variational
form of the problem may still seem rather abstract since it appears -- even in the dual -- that we need to
check an infinite number of values of $\mu$, for each choice of the vector, $\nu$.
However, it suffices in applications to consider a fine grid of values 
$\{ \mu_1 , \cdots , \mu_m \}$ and write the primal problem as
\[
\min_{f \in \RR^m, g \in \RR^n} \Big\{ - \sum_{i=1}^n \log(g_i) \; \Big| \; 
  Af = g, f \in \SS \Big\}
\]
where $A$ is an $n$ by $m$ matrix with elements $p ( x_i | \mu_j)$ and
$\SS = \{s \in \RR^m | 1^\top s = 1, \; s \geq 0 \}$  is the unit simplex.
Thus, $\hat f_j$ denotes the estimated mixing density evaluated at the grid point,
$\mu_j$ and $\hat g_i$ denotes the estimated mixture density evaluated at $x_i$.
The dual problem in this discrete formulation becomes,
\[
\max_{\nu \in \RR^n} \Big\{ \sum_{i=1}^n \log \nu_i \; \Big| \; 
A^\top \nu \leq n 1_m, \quad \nu \geq 0 \Big\}.
\]
Primal and dual solutions are immediately recoverable from the solution to either
problem.  Interior point methods such as those provided by PDCO of \citeasnoun{PDCO} and
Mosek of \citeasnoun{Mosek}, are capable of solving dual formulations of typical problems
with $n = 200$ and $m = 300$ in less than one second.
The empirical Bayes package {\tt REBayes},  \citeasnoun{REBayes}, is available 
for download from the R repository CRAN.  It is based on the {\tt RMosek} package of
\citeasnoun{RMosek}, and was used for all of the computations reported below.
We have compared this approach with other proposals including those of
\citeasnoun{LK} and \citeasnoun{GJW08}, but thus far have found nothing competitive 
in terms of speed and accuracy.

Solutions to the nonparametric MLE problem of Kiefer and Wolfowitz produce estimates
of the mixing measure, $\eta$, that are discrete and possess only a few mass points.
A theoretical upper bound  on the number of these atoms of $\eta$ was established already
by \citeasnoun{LindsayI}, but in practice the number is typically observed  to be far
fewer.  It may seem surprising, perhaps even disturbing, that even when the true mixing
distribution has a smooth density, the NPMLE estimate of that density is discrete with only
a few atoms.  However, this may appear less worrying if we consider a more explicit
example.  Suppose that we have a location mixture of Gaussians,
\[
g(x) = \int \phi (x - \mu) d\eta(\mu),
\]
so we are firmly in the deconvolution business, a harsh environment notorious for its
poor convergence rates.  One interpretation of this is that good approximations of the
mixture density $g$ can be achieved by relatively simple  discrete mixtures with only a 
few atoms.  For many applications estimation of $g$ is known to be sufficient:  this is quite
explicit for example for empirical Bayes compound decision problems where the Bayes rules
are known to depend entirely on the estimated $\hat g$.  See e.g. \citeasnoun{efron.11}.  
Of course given our discrete
formulation of the Kiefer-Wolfowitz problem, we can only identify the location of
atoms up to the scale of the grid spacing, but we believe that the $m \approx 300$
grid points we have been using in the simulations reported below 
are probably adequate for most applications. For testing this assertion is reinforced
by the fact that finer grids, when employed, exert a negligible impact on the LRT statistic. 
Recently, \citeasnoun{DickerZhao} have shown that with $m = \sqrt{n}$, the Hellinger distance 
between $\hat g$ and $g$ is bounded by $\mathcal{O}_{p}(\log n/\sqrt{n})$. 

Given a reliable maximum likelihood estimator for the general nonparametric mixture
model it is of obvious interest to know whether an effective likelihood ratio testing
strategy can be developed.  This question has received considerable prior attention,
again \citeasnoun{Lindsay95} provides an authoritative overview of this literature.
However, more recently work by \citeasnoun{gassiat2002} and \citeasnoun{Azgame2009} has 
revealed new features of the asymptotic behavior of the likelihood ratio for mixture
settings that enable one to derive asymptotic critical values for the LRT.

\subsection{Asymptotic Theory of Likelihood Ratios for General Mixtures}
Consider a parametric family of distributions that have density $p(\cdot|\mu)$ with respect to some sigma-finite measure $\lambda$ and parameters from the parameter set $\Theta \subset \R^d$. Our aim is to test whether the i.i.d. sample $X_1,...,X_n$ was generated from a $p(\cdot|\mu_0)$ for some $\mu_0 \in \Theta$ against the general alternative that $X_1,...,X_n$ is generated from a mixture of the form $p_\eta(\cdot) := \int_\Theta p(\cdot|\mu) d\eta(\mu)$ for some non-degenerate distribution $\eta$ on $\Theta$ (non-degenerate in the sense that $\eta$ is  not a one-point distribution). In order for this testing problem to make sense, we need the following mild identifiability assumption
\begin{enumerate}
\item[(A0)] For any probability measure $\eta$ on $\Theta$, for any $\mu_0 \in \Theta$ we have $\eta \neq \delta(\mu_0)$ (denoting by $\delta(\mu)$ the Dirac measure at the point $\mu$) implies $\E[(p_\eta(X_1) - p(X_1|\mu_0))^2] > 0$.
\end{enumerate}
Consider the following sets of distributions on $\Theta$
\[
\bar{\Gc} := \{\eta | \eta \mbox{ distribution on } \Theta, \}, \quad \Gc := \bar{\Gc} \backslash \delta(\mu_0).
\]
Define the log-likelihood function corresponding to the measure $\eta$ as
\[
\ell_n(\eta) := \sum_{i=1}^n \log p_\eta(X_i).
\]
The likelihood ratio test statistic is given by 
\[
L_n := \sup_{\eta\in\bar\Gc} \ell_n(\eta ) - \sup_{\mu \in \Theta} \ell_n(\delta(\mu)).
\]
To derive the asymptotic distribution of the likelihood ratio under the null, assume that the data are generated from a measure with density $p(\cdot|\mu_0)$ for some $\mu_0 \in \Theta$. Consider the decomposition
\[
L_n = \sup_{\eta \in \bar\Gc} \ell_n(\eta) - \ell_n(\delta(\mu_0)) + \ell_n(\delta(\mu_0)) - \sup_{\mu \in \Theta} \ell_n(\delta(\mu)).
\] 
The second term in this decomposition can be handled by classical parametric theory. Under suitable regularity conditions we obtain
\begin{equation}\label{eq:classlr1}
\sup_{\mu \in \Theta} \ell_n(\delta(\mu)) - \ell_n(\delta(\mu_0)) = \frac{1}{2}\Big\| \frac{1}{\sqrt{n}} \sum_{i=1}^{n} I(\mu_0)^{-1/2}\ell'(X_i|\mu_0)\Big\|^2 + o_P(1) 
\end{equation} 
with $\ell'(X_i|\mu) := \nabla_{\mu} \log p_{\delta(\mu)}(X_i)$, and 
$I(\mu_0) = \E[\ell'(X_i|\mu_0)\ell'(X_i|\mu_0)^\top]$ being the Fisher information.
Handling the first part in the decomposition is more challenging. Expansions for this term were derived in \cite{gassiat2002,liushao2003,Azgame2009} under various sets of conditions. For the sake of a simple presentation we will follow \citeasnoun{gassiat2002}. For $\eta \in \bar \Gc, \mu \in \Theta, \eta \neq \delta(\mu)$ let
\begin{equation} \label{eq:score}
s_{\eta,\mu}(x) := \Big(\frac{p_\eta(x)}{p_{\delta(\mu)}(x)} -1  \Big) \Big/ \Big\|\frac{p_\eta}{p_{\delta(\mu)}} -1 \Big\|_{2,\delta(\mu)}
\end{equation}
where we defined $\|f\|_{2,\eta} := (\int \int f^2(x)p(x|z) d\eta(z)d\lambda(x))^{1/2}$. For $\eta \in \Gc$ define 
\[
\Gb_n(\eta) := n^{-1/2}\sum_{i=1}^n s_{\eta,\mu_0}(X_i)
\]
and note that by construction $\E[s_{\eta,\mu_0}(X_i)] = 0, \E[s_{\eta,\mu_0}^2(X_i)] = 1$. Now a slight modification of the proof of Theorem 3.1 in \citeasnoun{gassiat2002} leads to the following result for the asymptotic behavior of the likelihood ratio test - for the sake of completeness a sketch of the proof is provided in the Appendix. 

\begin{theorem}\label{th:iid}
Assume $X_1,...,X_n$ are generated from $p(\cdot|\mu_0)$, that (A0) holds and that $\Gb_n \weak \Gb$ in $\ell^\infty(\Gc)$ for a centered Gaussian process $\Gb$. 
Then 
\begin{equation} \label{eq:iid1}
2 \left(\underset{\eta \in \bar\Gc}{\sup} \ell_{n}(\eta) - \ell_{n}(\delta(\mu_0))\right ) = \underset{\eta \in \mathcal{G}}{\sup} \Big ( \max \Big \{ \mathbb{G}_n(\eta), 0 \Big \} \Big)^2 + o_P(1). 
\end{equation}
If additionally \eqref{eq:classlr1} holds and $\ell'(X_1|\mu_0)$ is square integrable, 
\[
 2 L_{n} \weak \underset{\eta \in \mathcal{G}}{\sup} \Big ( \max \Big \{ \mathbb{G}(\eta), 0 \Big \} \Big)^2 - \|Y\|^{2}. 
\]
Here, $Y \sim \mathcal{N}(0,I_d)$ and $(\mathbb{G}, Y)$ is jointly centered normal with covariance taking the form $\E[\mathbb{G}(\eta) Y] = \E[s_{\eta, \mu_0}(X_1) I(\mu_0)^{-1/2} \ell'(X_1|\mu_0)], \Cov(\mathbb{G}(\zeta),\mathbb{G}(\eta)) = \E[s_{\zeta, \mu_0}(X_1) s_{\eta, \mu_0}(X_1)]$. Here, by jointly normal we mean that for any collection $\eta_1,...,\eta_k \in \Gc$ the vector $(Y_1,\Gb(\eta_1), ...,\Gb(\eta_k))$ follows a centered multivariate normal distribution with the covariance described above.
\end{theorem}

\subsection{Asymptotic Critical Values}
In order to apply the above limiting result in practice, we need to know how to obtain critical values from the asymptotic distribution. For illustrative purposes, we consider the following normal mixture example. 

\begin{example} \label{eg:cval}
Consider mixtures of $\Nc(\mu,1)$ distributions and assume that $\Theta = [L,U]$ with $0 \in \Theta$. Computations in \citeasnoun{Azgame2009} show that the asymptotic distribution of the log-likelihood ratio test statistic $L_{n}$ under the null of $X_i \sim \Nc(0,1)$ i.i.d. is given by
\[
D = \Big( \sup_{\eta\in \mathcal{G}} (V_\eta)_+\Big)^2 - Y_1^2
\]
where $(V_\eta)_{\eta \in \mathcal{G}}$ is the Gaussian process given by
\[
V_\eta := \Big(\sum_{k=1}^\infty \frac{Y_k \kappa_k(\eta)}{(k!)^{1/2}}\Big)\Big/\Big(\sum_{k=1}^\infty \frac{\kappa_k^2(\eta)}{k!}\Big)^{1/2}
\]
with $Y_1,Y_2,...$ denoting i.i.d. $\Nc(0,1)$ distributed random variables, $\kappa_k(\eta) := \int_\Theta \mu^k d\eta(\mu)$ and $x_+$ denoting the positive part of $x$. \\

There exists a simpler expression for the distribution of $D$. More precisely, we will demonstrate that
\begin{equation} \label{approx}
D \deq \sup_{\eta \in \mathcal{G}}\Big(\Big( \Big(\sum_{k=2}^\infty \frac{Y_k \kappa_k(\eta)}{(k!)^{1/2}}\Big)_+\Big)^2\Big/\sum_{k=2}^\infty \frac{\kappa_k^2(\eta)}{k!}\Big).
\end{equation}
The detailed derivation is provided in the Appendix. Approximating the distribution function of the measure $\eta$ on $\Theta$ by a discrete distribution function with masses $p_1,...,p_N$ on a fine grid $m_1,...,m_N$ leads to the approximation
\[
D \approx \sup_{p_1,...,p_N}\Big(\Big(  \Big(\sum_{j=1}^N p_j\sum_{k=2}^\infty \frac{Y_k m_j^k}{(k!)^{1/2}}\Big)_+\Big)^2\Big/\sum_{i,j=1}^N p_ip_j\sum_{k=2}^\infty \frac{(m_j m_i)^k}{k!}\Big).
\]
In particular, maximizing the right-hand side with respect to $p_1,...,p_N$ under the constraints $p_i \geq 0, \sum p_i = 1$ for fixed grid $m_1,...,m_N$ can be formulated as a quadratic optimization problem of the form
\[
\min_p p^\top  A p \quad \mbox{under} \quad p_i \geq 0,\  p^\top b = 1
\]
where $p=(p_1,...,p_N)$, $A_{ij} = \sum_{k=2}^\infty \frac{(m_j m_i)^k}{k!}$, $b_i = \sum_{k=2}^\infty \frac{Y_k m_i^k}{(k!)^{1/2}}$, if $\underset{i}{\max}~b_i >0$. If $\underset{i}{\max}~b_i \leq 0$, we can set $D=0$. This suggests a practical way of simulating critical values after replacing the infinite sum by a finite approximation and avoiding the grid point $0$. Table~\ref{table1} below contains simulated critical values in some particular settings. All results are based on $10,000$ simulation runs with the sums for $A$ and $b$ cut off at $k=25$ and grids with $200$ points equally spaced points excluding the point $0$.
\end{example}

\begin{table}[ht]
\center{
\begin{tabular}[c]{c|ccc}
\hline
$\Theta$ & $90\%$ & $95\%$ & $99\%$ \\
\hline
[-1,1]& 2.75 & 3.95 & 6.93 \\
\hline
[-2,2]& 3.90 & 5.37 & 8.71 \\
\hline
[-3,3]& 5.34 & 6.87 & 10.46 \\
\hline
[-4,4]& 6.38 & 8.32 & 11.91 \\
\hline
\end{tabular}}
\vspace{3mm}
\caption{Simulated asymptotic critical values for the asymptotic null distribution for various choices
of the set $\Theta$.}
\label{table1}
\end{table}

To explore the finite sample performance of the above method we begin with an
experiment to compare the critical values of the LRT of homogeneity in the Gaussian location
model with the simulated asymptotic critical values in Table~\ref{table1}.  We consider sample sizes, 
$n \in \{ 100, 500, 1000, 5000, 10000 \}$ and four choices of the domain of the 
MLE of the mixture are considered:  $\{ [-j,j]: j = 1,\cdots,4  \}$.  
We maintain a grid spacing of 0.01 for the mixing distribution on these domains for each of
these cases for the Kiefer-Wolfowitz MLE.  Results are reported in Table 2.  
For the three largest sample sizes we bin the observations into 300 and 500 equally
spaced bins respectively.  It will be
noted that the empirical critical values are consistently smaller than those simulated
from the asymptotic theory.  There appears to be a tendency for the empirical 
critical values to increase with $n$, but this tendency is rather weak.  This finding is perhaps
not entirely surprising in view of the slow rates of convergence established elsewhere in the
literature, see e.g. \citeasnoun{BickelChernoff} and \citeasnoun{HallStewart}. 
These findings imply that our simulated asymptotic critical values are not likely to work well 
for size control, which motivates us to consider an alternative bootstrap based method in 
determining critical values in the next section.

{\small{ 
\begin{table}[!tbp]
\begin{center}
\begin{tabular}{rrrrrcrrrrcrrrr}
\hline\hline
\multicolumn{1}{c}{\bfseries n}&\multicolumn{4}{c}{\bfseries cval(.90)}&\multicolumn{1}{c}{\bfseries }&\multicolumn{4}{c}{\bfseries cval(.95)}&\multicolumn{1}{c}{\bfseries }&\multicolumn{4}{c}{\bfseries cval(.99)}\tabularnewline
\cline{2-5} \cline{7-10} \cline{12-15}
\multicolumn{1}{r}{}&\multicolumn{1}{c}{[-1,1]}&\multicolumn{1}{c}{[-2,2]}&\multicolumn{1}{c}{[-3,3]}&\multicolumn{1}{c}{[-4,4]}&\multicolumn{1}{c}{}&\multicolumn{1}{c}{[-1,1]}&\multicolumn{1}{c}{[-2,2]}&\multicolumn{1}{c}{[-3,3]}&\multicolumn{1}{c}{[-4,4]}&\multicolumn{1}{c}{}&\multicolumn{1}{c}{[-1,1]}&\multicolumn{1}{c}{[-2,2]}&\multicolumn{1}{c}{[-3,3]}&\multicolumn{1}{c}{[-4,4]}\tabularnewline
\hline
100&$2.09$&$2.69$&$2.80$&$2.80$&&$3.07$&$3.70$&$3.97$&$4.06$&&$6.43$&$7.58$&$ 8.31$&$ 8.55$\tabularnewline
500&$2.22$&$2.80$&$2.96$&$2.98$&&$3.06$&$3.87$&$4.41$&$4.41$&&$5.69$&$7.07$&$ 7.45$&$ 7.52$\tabularnewline
1,000&$2.67$&$3.46$&$3.72$&$3.76$&&$3.73$&$4.95$&$5.44$&$5.56$&&$7.26$&$8.55$&$ 9.51$&$ 9.76$\tabularnewline
5,000&$2.68$&$3.56$&$3.91$&$3.96$&&$3.79$&$4.54$&$4.83$&$5.09$&&$6.52$&$8.15$&$ 8.32$&$ 8.38$\tabularnewline
10,000&$2.41$&$3.11$&$3.29$&$3.46$&&$3.61$&$4.45$&$4.72$&$4.97$&&$6.23$&$7.51$&$ 7.96$&$ 8.32$\tabularnewline
$\infty$&$2.75$&$3.90$&$5.34$&$6.38$&&$3.95$&$5.37$&$6.87$&$8.32$&&$6.93$&$8.71$&$10.46$&$11.91$\tabularnewline
\hline
\end{tabular}
\end{center}
\caption{Critical Values for Likelihood Ratio Test of Gaussian Parameter Homogeneity:  
The first five rows of the table report empirical critical values based on 1000 
replications of the LRT based on the Kiefer-Wolfowitz estimate of the nonparametric 
Gaussian location mixture distribution.  Results for sample sizes 5,000 and 10,000 
were computed by binning the observations into 300, 500 equally spaced bins 
respectively.
Restriction of the domain of the mixing distribution is indicated by the column labels.
The last row reproduces the simulated asymptotic critical values reported in Table 1.\label{B}} 
\end{table}

 }}

\subsection{A Parametric Bootstrap Method for Critical Values} \label{sec: bcval}
The parametric bootstrap method for testing parameter homogeneity we are about to introduce is a very natural idea. In finite mixture models, similar approaches have been proposed by \citeasnoun{McLachlan87} and \citeasnoun{chenchen01}. However, to the best of our knowledge, this is the first time that such a bootstrap method has been formally shown to produce consistent critical values for likelihood ratio tests in mixture models. \\

The parametric bootstrap approach to determine critical values for the distribution of $L_n$ is defined as follows.
\begin{enumerate}
\item Compute the maximum likelihood estimator $\hat \mu := \mbox{argmax}_{\mu \in \Theta} \ell_n(\delta(\mu))$.
\item For $b=1,...,B$ generate data $X_{1,n}^{(b)},...,X_{n,n}^{(b)} \sim p(\cdot|\hat \mu)$ i.i.d. 
\item For $b=1,...,B$ denote by $L_{n,b}$ the statistic $L_n$ computed from the sample $X_{1,n}^{(b)},...,X_{n,n}^{(b)}$. Compute the $\alpha$-quantile $q_{n,\alpha}$ of $L_{n,1},...,L_{n,B}$.  
\end{enumerate}
The null of parameter homogeneity is rejected if $L_n > q_{n,1-\alpha}$. To prove that this bootstrap 
procedure leads to a valid (asymptotic) test, we need to show that $P(L_n > q_{n,1-\alpha}) \to \alpha$ 
if $X_1,...,X_n$ are generated under the null. To establish this result, we need two main ingredients. 
First, we need to analyze the limiting properties of the likelihood ratio test for data that are generated 
under triangular arrays. This is done in Theorem~\ref{th:main}. Second, we need to establish continuity 
of the limiting distribution of $F_R$ around its $\alpha-$quantile. This is done in Theorem~\ref{th:continuity}.  
Together, Theorem \ref{th:main} and \ref{th:continuity} imply consistency of the proposed bootstrap procedure.

We now require some additional notation. Fix an arbitrary sequence of points $\mu_n$ in $\Theta \subset \R^d$ with $\mu_n \to \mu_0 \in \Theta$ as $n \to \infty$. For $\eps > 0$, define $\Theta^\eps$ as the $\eps$-enlargement of $\Theta$ with respect to Euclidean distance. Let
\[
\bar{\Gc}^\eps := \{\eta | \eta \mbox{ distribution on } \Theta^\eps \}, \quad \Gc^\eps := \bar{\Gc}^\eps \backslash \delta(\mu_0).
\]
To each measure $\eta \in \Gc$ define the measure $\eta_n$ through $\eta_n(A) = \eta(A - \mu_n + \mu_0)$ for all Borel sets $A \subset \Theta$ where $A + x :=\{a+x|a\in A\}$ for a set $A \subset \R$ and $x \in \R$. From now on, assume that $X_{1,n},...,X_{n,n}$ are i.i.d. $\sim p(\cdot|\mu_n)$ and consider the following sequence of processes indexed by $\Gc^\eps$ 
\[
\Gb_{n}^*(\eta) := n^{-1/2} \sum_{i=1}^n s_{\eta_n,\mu_n}(X_{i,n})
\]  
where the scores $s_{\eta_n,\mu_n}$ are defined in \eqref{eq:score}. Write $\ell_n^*(\eta) := \sum_{i=1}^n \log p_\eta(X_{i,n})$. To analyze the asymptotic behavior of $L_n^* := \sup_{\eta \in \bar{\Gc}} \ell_n^*(\eta) - \sup_{\mu \in \Theta} \ell_n^*(\delta(\mu))$, consider the decomposition
\[
L_n^* = \sup_{\eta \in \bar{\Gc}} \ell_n^*(\eta) - \ell_n^*(\delta(\mu_n)) + \ell_n^*(\delta(\mu_n)) - \sup_{\mu \in \Theta} \ell_n^*(\delta(\mu)).
\]
Classical results suggest that under suitable regularity conditions the second part in the above decomposition should take the form
\begin{equation}\label{eq:classlr2}
\sup_{\mu \in \Theta} \ell_n^*(\delta(\mu)) - \ell_n^*(\delta(\mu_n)) = \frac{1}{2}\Big\| \frac{1}{\sqrt{n}} \sum_{i=1}^{n} I(\mu_n)^{-1/2}\ell'(X_{i,n}|\mu_n)\Big\|^2 + o_P(1) 
\end{equation} 
provided that $\mu_n \to \mu_0$. Various conditions ensuring the above representation exist, and we are not going into details here. The main challenge is to derive an expansion for the first part of $L_n^*$. Such an expansion is established in Theorem~\ref{th:main} under the following set of assumptions:
\begin{enumerate}
\item[(A1)]
Assume that 
\[
\Big(\Gb_{n}^*,\frac{1}{\sqrt{n}} \sum_{i=1}^{n} I(\mu_n)^{-1/2}\ell'(X_{i,n}|\mu_n)\Big) \weak (\Gb^*, Y_1)
\] 
in $\ell^\infty(\Gc^\eps)\times \R$ where $(\Gb^*, Y)$ are jointly centered normal with $Y \sim \Nc(0,I_d)$ and covariance structure of the form, 
\begin{align*}
\E [\Gb^*(\eta_1)\Gb^*(\eta_2)] &= \int_\R s_{\eta_1,\mu_0}(x)s_{\eta_2,\mu_0}(x)p_{\delta(\mu_0)}(x)d\lambda(x),
\\
\E[\Gb^*(\eta) Y_1] &= \int_\R s_{\eta, \mu_0}(x) I(\mu_0)^{-1/2} \ell'(x|\mu)p_{\delta(\mu_0)}(x)d\lambda(x).
\end{align*}
Additionally, assume that for $\eps \downarrow 0$ we have
\begin{equation} \label{eq:epsto0*}
\sup_{\eta \in \Gc^\eps} \inf_{\tilde \eta \in \Gc} |\Gb^*(\eta) - \Gb^*(\tilde\eta)|   = o_P(1).
\end{equation}
\item[(A2)] Letting $s_{\eta,\mu,-} := \min\{0,-s_{\eta,\mu}\}$ we have that 
\[
\sup_{\eta \in \Gc^\eps} \Big|\frac{1}{n}\sum_{i=1}^n (s_{\eta_n,\mu_n}^2(X_{i,n}) - 1)\Big| + \Big|\frac{1}{n}\sum_{i=1}^n (s_{\eta_n,\mu_n,-}^2(X_{i,n}) - \|s_{\eta,\mu_0,-}\|^2_{2,\delta(\mu_0)})\Big|    = o_P(1). 
\]
\item[(A3)] For every $n \in \N$, assume that the class of functions
\[
\Fc_n := \Big\{ x \mapsto s_{\eta,\mu_n}(x) \Big| \eta \in \Gc \Big\}
\]
admits an envelope function $F_n$ such that $\max_{i=1,...,n} F_n(X_{i,n}) = o_P(n^{1/2})$. 
\end{enumerate}

\begin{remark}
Note that the process $\Gb_n^*$ is indexed by measures $\eta$, and not by the score functions $s_{\eta_n,\mu_n}$ where the latter would correspond to 'classical' empirical process theory. The reason for this indexing is that the score functions $s_{\eta_n,\mu_n}$ depend on $n$. Thus indexing by score functions $s_{\eta_n,\mu_n}$ we would obtain an index set which depends on $n$, which would lead to various technical problems. On the other hand, using $s_{\eta_n,\mu_n}$ instead of $s_{\eta,\mu_n}$ in the definition of $\Gb^*_n$ is crucial since $s_{\eta,\mu_n}$ can be quite different for the same values of $\eta$ but different $\mu_n$. As an example of the latter, let $\mu_n = \mu_0 + 1/n$, $\tilde \mu_n = \mu_0 + 3/n,$ $\eta = \delta(\mu_0 + \alpha)$. Then, for $\alpha$ small, under suitable differentiability conditions we have $s_{\eta,\delta(\mu_n)}(x) \approx sgn(\alpha - 1/n) \ell'(x|\mu_n)/\|\ell'(x|\mu_n)\|_{2,\delta(\mu_n)}$ and $s_{\eta,\delta(\tilde\mu_n)}(x) \approx sgn(\alpha - 3/n) \ell'(x|\tilde\mu_n)/\|\ell'(x|\tilde\mu_n)\|_{2,\delta(\tilde\mu_n)}$. For $\alpha \in (1/n,3/n)$ the sign of $\alpha - 1/n$ and $\alpha - 3/n$ will differ, and this leads to different score functions. This problem does not arise if we use $s_{\eta_n,\mu_n}$ instead. \hfill $\blacksquare$  
\end{remark}

For location-shift mixtures, that is mixtures of densities of the form $p(\cdot|\mu) = p(\cdot - \mu)$, assumptions (A1)-(A3) can be considerably simplified.

\begin{prop}\label{prop:loc-scale}
Assume that $p(\cdot|\mu) = p(\cdot - \mu)$, the conditions of Theorem~\ref{th:iid} hold with $\Gc^\eps$ instead of $\Gc$, that \eqref{eq:classlr2} holds, and that additionally for $\gamma \downarrow 0$ we have, for $\Gb$ denoting the weak limit of $\Gb_n$ in Theorem~\ref{th:iid}, 
\begin{equation} \label{eq:epsto0}
\sup_{\eta \in \Gc^\gamma} \inf_{\tilde \eta \in \Gc} |\Gb(\eta) - \Gb(\tilde\eta)|   = o_P(1).
\end{equation}
Then conditions (A1)-(A3) hold.
\end{prop}



The proof of Proposition~\ref{prop:loc-scale} repeatedly makes use of the fact that the assumptions of Theorem~\ref{th:iid} hold for $\Gc^\eps$ instead of $\Gc$. In general, this can not be avoided. Intuitively, this is due to the fact that for measures $\eta$ with support in $\Theta$ the support of $\eta_n$ will not necessarily be contained in $\Theta$.

Next, we show that assumptions (A1)-(A3) are realistic and can be verified for some standard models. 

\begin{example} (Location Mixture of Gaussians) \label{ex:gauss}
Assume that $\Theta = [a,b]$ for some $a<0<b$ and that the densities $p$ take the form $f(x|\mu) = (2\pi)^{-1/2}\exp((x-\mu)^2/2)$. Without loss of generality we will assume that $\mu_0 = 0$. In this setting, the densities have the location-scale structure described in Proposition~\ref{prop:loc-scale}, and thus it suffices to verify the conditions of  Theorem~\ref{th:iid} hold with $\Gc^\eps$ instead of $\Gc$, that \eqref{eq:classlr2} holds, and that \eqref{eq:epsto0} is satisfied. Note that~\eqref{eq:classlr2} can be established by standard arguments, the details are omitted for the sake of brevity. 

The arguments from the proof of Theorem 3 in \cite{Azgame2009} yield $\Gb_n \weak \Gb$ in $\ell^\infty(\Gc^\eps)$ where the limiting process $\Gb$ is Gaussian and has a covariance structure of the form
\[
\E[\Gb(\eta_1)\Gb(\eta_2)] = \frac{\E[\exp(Z_1Z_2)] - 1}{(\E[\exp(Z_1\tilde Z_1)] - 1)^{1/2}(\E [\exp(Z_2\tilde Z_2)] - 1)^{1/2}} 
\]  
where $Z_1,\tilde Z_1 \sim\eta_1, Z_2,\tilde Z_2 \sim \eta_2$ and $Z_1,Z_2,\tilde Z_1, \tilde Z_2$ are independent. Joint asymptotic normality with $Y_1$ follows by standard arguments. To prove~\eqref{eq:epsto0}, consider the following construction. To each random variable $Z$ on $[a-\eps,b+\eps]$ define a transformed random variable $W$ through 
\[
W := ZI\{Z \in [a,b]\} + \frac{M}{M+\eps}ZI\{Z \notin [a,b]\}.
\]
where $M := \min(|a|,b)$. By construction, the support of $W$ is contained in $[a,b]$. Denoting the distribution of $W$ by $\xi_{\eta,\eps}$, straightforward but tedious calculations show that
\[
\sup_{\eta \in \Gc^{\eps}} \E[(\Gb(\eta) - \Gb(\xi_{\eta,\eps}))^2]= o(1)
\]
as $\eps \downarrow 0$. By the uniform continuity of the process $\Gb$ with respect to the metric $d(\eta,\xi) := (\E[(\Gb(\eta) - \Gb(\xi))^2])^{1/2}$ induced by its covariance [see Example 1.5.10 in \cite{vandwell1996}], this shows that~\eqref{eq:epsto0} also holds.
\hfill $\blacksquare$  
\end{example}

\begin{remark}\label{rem:gausslocscale}
As pointed out by a Referee, location-scale mixtures on Gaussians, i.e. mixtures of the form $p(x|\eta) = \iint p(x|\mu,\sigma) d\eta(\mu,\sigma)$ with $p(\cdot|\mu,\sigma)$ denoting the density of an $\Nc(\mu,\sigma^2)$ random variable, are also of practical interest.  In such models, even identification of parameters is a very subtle issue. To illustrate this point, consider a location mixture of normals with unknown variance parameter. If the support of the location parameter is unrestricted,  assumption (A0) will fail if we allow for general classes of mixtures. 
To see that, denote by $\eta(\tau)$ the product of an $\Nc(0,\sigma^2-\tau^2)$ measure for location and a point mass at $\tau^2$ for variance where $0 \leq \tau^2 \leq \sigma^2$. Then $p_{\eta(\tau)} \equiv p_{\eta(\tau')}$ for any $\tau,\tau' \in [0 ,\sigma]$, and setting $\tau^2 = \sigma^2$ corresponds to homogeneity. Thus (A0) does not hold. 
Assuming that the support for $\mu$ is restricted to a compact set, the unknown variance $\sigma^2$ and the mixing distribution can be jointly identified. We are not aware of results on identification if both, location and scale are being mixed, even if the support for both parameters is confined to compact sets. Gaining a better understanding of identification and, provided identification holds, the behaviour of LRT in this case is a very interesting and important question. We leave this question to future research.  
 \hfill $\blacksquare$
\end{remark}

\begin{example} (Mixture of Poisson distributions)
Assume that $\Theta = [a,b]$ for some $0<a<b$ and that the densities $p$ take the form $p(k|\mu) = \mu^ke^{-\mu}/k!$ with respect to the counting measure on $\N$. Note that this model does not have the location-scale structure discussed in Proposition~\ref{prop:loc-scale}. Assumptions (A1)-(A3) can still be verified, and the technical details are provided in Section~\ref{sec:poisdet} of the Appendix.  \hfill $\blacksquare$  
\end{example}

We now state our main result.

\begin{theorem} \label{th:main}
Under assumptions (A0)-(A3) we have
\begin{equation} \label{eq:main1}
2\sup_{\eta \in \bar \Gc}\Big(\ell_n^*(\eta) - \ell_n^*(\delta(\mu_n))\Big)  = \sup_{\eta \in \Gc} \Big(\max\Big\{\Gb_n^*(\eta), 0\Big\}\Big)^2 + o_P(1).
\end{equation}
If additionally \eqref{eq:classlr2} holds we have
\[
2\Big(\sup_{\eta \in \bar \Gc }\ell_n^*(\eta) - \sup_{\mu \in \Theta} \ell_n^*(\delta(\mu))\Big) \weak R := \sup_{\eta \in \Gc} \Big(\max(\Gb^*(\eta),0)\Big)^2 - Y_1^2.
\]
\end{theorem}

Intuitively, Theorem \ref{th:main} suggests that critical values based on the parametric bootstrap 
should lead to an asymptotic level $\alpha$ test of homogeneity. However, a formal proof of this 
statement requires that the distribution of $R$, say $F_R$, is continuous at $F_R^{-1}(\alpha)$. 
The following theorem completes this last step. 

\begin{theorem} \label{th:continuity}
Let the assumptions of Theorem \ref{th:iid} hold. Then the distribution of $R$ is continuous on $(0,\infty)$ and $P(R<0)=0$. Provided that $B = B_n \to \infty$ we have $\limsup_{n \to \infty} P(L_n > q_{n,1-\alpha}) = \alpha$ for any $\alpha$ satisfying $P(R>0) > \alpha$. Moreover, if $d=1$ and if there exists $\eta \in \Gc$ such that $\E[\Gb(\eta) Y_1] \neq \pm 1$ we have $P(R>0) \geq 1/4$.
\end{theorem}


\begin{remark} \label{remark:support} 
How to choose support to solve for the NPMLE is a very important practical question. For location shift models, it is easy to show that the NPMLE $\hat\eta$ will not have any mass points outside of the sample support. This type of result has been generalized in \citeasnoun{Lindsay81} to other univariate base densities that have a unique mode. In particular, suppose that for each sample point $x_i$, the function $\mu \mapsto p(x_i\mid \mu)$ has a unique mode at $\mu_i^*$. Then the support od the NPMLE $\hat\eta$ must be contained in $[\mu_{m}^{*}, \mu_{M}^*]$ where $\mu_m^*$ and $\mu_M^*$ are the minimum and maximum of $(\mu_i^*, \dots, \mu_n^*)$, respectively. This is true for many base distributions in the exponential family. For example, for mixtures of exponential distributions with mean $\exp(-\phi)$, the mode for the base density $\exp(\phi) \exp(-x_i\exp(-\phi))$ is located at $\phi_i^* = -\ln (x_i)$. Hence the support for the mixture distribution must be contained in $[\min (\ln(1/x_i)), \max(\ln(1/x_i))]$. To ensure compactness of the parameter space, we recommend taking the 5-th and 95-th quantile of $\mu_1^*,...,\mu_n^*$.
\end{remark}

\begin{remark} \label{remark:invariance} 
For mixture models with densities of the form $p(\cdot|\mu) = p(\cdot-\mu)$ there is an alternative way of simulating quantiles of the LR test. The key observation is that, assuming that we allow for an arbitrary support of the mixing distribution, the distribution of the likelihood ratio test under the null does not depend on the location of the true parameter. More precisely, assume that $X_1,...,X_n$ generated from $p(\cdot|\mu_X)$ and $Y_1,...,Y_n$ are generated from $p(\cdot|\mu_Y)$. Then $X_i$ has the same distribution as $Y_i - \mu_Y + \mu_X$, and for any measure $\eta$ the log-likelihood $\sum_{i=1}^n \log p_\eta(X_i)$ has the same distribution as $\sum_{i=1}^n \log p_\eta(Y_i  - \mu_Y + \mu_X )$, which equals the distribution of $\sum_{i=1}^n \log p_{\tilde\eta}(Y_i)$ with the measure $\tilde \eta$ defined through $\tilde \eta(A) = \eta(A - \mu_Y + \mu_X )$. This implies that the log-likelihood ratio test statistic computed from $X_1,...,X_n$ and the one computed $Y_1,...,Y_n$ will have the same distribution.

Thus the following procedure provides a way to conduct an exact test for parameter homogeneity when the support of the mixing distribution is unrestricted.
\begin{enumerate}
\item Repeatedly generate data $Y_{1}, \dots, Y_{n} \sim p(\cdot|0)$ i.i.d. for $B$ times. For each bootstrap sample, compute the LR test statistics $L_{n, b}$ for $b = 1, \dots, B$. 
\item Compute the $1-\alpha$-quantile $q_{n,1-\alpha}^{L}$ of the bootstrap sample $L_{n,b}$, $b = 1, \dots, B$. 
\end{enumerate}
The null of parameter homogeneity is rejected if $L_{n} > q_{n,1-\alpha}^{L}$. \\

Table \ref{tab:bcval} tabulates the bootstrap critical values for the null distribution of the LR test statistics for testing homogeneity of the Gaussian location parameter. $B$ bootstrap samples of size $n$ is generated from standard normal distribution and the critical values are found based on the empirical distribution of the corresponding likelihood ratio test statistics.

\begin{table}[!h]
\begin{center}
\begin{tabular}{lrcrcr}
\hline\hline
\multicolumn{1}{l}{}&\multicolumn{1}{c}{90\%}&\multicolumn{1}{c}{}&\multicolumn{1}{c}{95\%}&\multicolumn{1}{c}{}&\multicolumn{1}{c}{99\%}\tabularnewline
\hline
n=100&$3.14$&&$4.60$&&$8.12$\tabularnewline
n=200&$3.15$&&$4.48$&&$ 7.21$\tabularnewline
n=500&$3.44$&&$4.69$&&$ 7.84$\tabularnewline
\hline
\end{tabular}
\end{center}
\caption{Bootstrap Critical Values for Likelihood Ratio Test of Homogeneity of Gaussian Location Parameter: For various sample sizes, the bootstrap critical values are found following the procedure described in Remark \ref{remark:invariance} with $B= 2,000$.} 
\label{tab:bcval}
\end{table}



It is important to keep in mind that this invariance property will hold only if we consider an 
\textit{unrestricted} support. In the case of Gaussian location mixtures, it is well known that the 
likelihood ratio test statistic with mixing distributions of unbounded support diverges to infinity 
(see \citeasnoun{Hartigan}). A more detailed analysis of this issue for some special cases of 
likelihood ratio tests in mixture models can be found in \citeasnoun{AGM2006} and \citeasnoun{HallStewart}. 
That analysis indicates that likelihood ratio test with unrestricted support can only detect local 
alternatives at slower rates than moment-based tests. However, the corresponding difference in rates 
is quite small and we compare via simulations the differences in power for using the parametric 
bootstrap critical values and the exact critical values for the location parameters in the Gaussian models. 
Results are summarized in Table~\ref{tab:parapower}, the power loss for reasonable sample sizes is 
quite modest.

To evaluate size performance of using these bootstrap critical values, we apply the LRT on a random 
sample $X_{1}, \dots, X_{n} \sim \mathcal{N}(1,1)$ for homogeneity versus general mixture on the 
location parameter. The third row of Table \ref{tab:sizenorm} reports the size performance of the 
LRT with these tabulated bootstrap critical values. In the same table, we also report the size 
performance of the LRT using critical values generated from the parametric bootstrap method, 
the $C(\alpha)$ test and the EM test that will be discussed in the next section. \hfill $\blacksquare$  
\end{remark}

\begin{table}[!tbp]
\begin{center}
\begin{tabular}{llcccc}
\hline\hline
\multicolumn{2}{l}{}&\multicolumn{1}{c}{90\%}&\multicolumn{1}{c}{95\%}&\multicolumn{1}{c}{99\%}\tabularnewline
\hline
\multirow{3}{*}{$h = 0.1$} & LRT-PBS[-1,1] &$0.2095$&$0.1180$&$0.0380$\tabularnewline
 &LRT-PBS[-2,2] &$0.2070$&$0.1135$&$0.0355$\tabularnewline
&LRT-EXT &$0.1765$&$0.1070$&$0.0375$\tabularnewline
\hline
\multirow{3}{*}{$h = 0.2$} & LRT-PBS[-1,1] &$0.6520$&$0.5120$&$0.2960$\tabularnewline
& LRT-PBS[-2,2]&$0.6255$&$0.4945$&$0.2550$\tabularnewline
& LRT-EXT&$0.5690$&$0.4505$&$0.2400$\tabularnewline
\hline
\multirow{3}{*}{$h = 0.3$} & LRT-PBS[-1,1] &$0.9775$&$0.9615$&$0.8805$\tabularnewline
& LRT-PBS[-2,2]&$0.9730$&$0.9485$&$0.8550$\tabularnewline
& LRT-EXT&$0.9660$&$0.9305$&$0.8430$\tabularnewline
\hline\hline
\end{tabular}\end{center}
\caption{Power comparison between parametric bootstrap method (denoted as LRT-PBS with stated support used for estimating the general mixture model) on restricted support and the Gaussian LRT with unrestricted support and exact critical value (denoted as LRT-EXT) as tabulated in Table~\ref{tab:bcval}. Simulation data is generated as $X_{1}, \dots, X_{n} \sim \mathcal{N}(\mu, 1)$ with $n = 200$ and $F_{\mu} = \frac{2}{3} \delta_{1.5h} + \frac{1}{3} \delta_{-3h}$ for $h$ taking values from $\{0.1, 0.2, 0.3\}$. Results are based on 2000 repetitions and the parametric bootstrap method is based on 500 bootstrap repetition on the stated support.} 
\label{tab:parapower}
\end{table}

\begin{table}[!h]
\begin{center}
\begin{tabular}{lrrrclrrrclrrr}
\hline\hline
\multicolumn{1}{l}{}&\multicolumn{3}{c}{$n=100$}&\multicolumn{1}{c}{\bfseries }&\multicolumn{3}{c}{$n=200$}&\multicolumn{1}{c}{\bfseries}&\multicolumn{3}{c}{$n=500$}\tabularnewline
& $90\%$ & $95\%$ & $99\%$ & &$90\%$ & $95\%$ & $99\%$  & & $90\%$ & $95\%$ & $99\%$  \tabularnewline \hline
EM &$0.088$&$0.044$&$0.010$&&$0.094$&$0.050$&$0.012$&&$0.094$&$0.048$&$0.010$\tabularnewline
$C(\alpha)$ &$0.103$&$0.050$&$0.018$&&$0.104$&$0.058$&$0.014$&&$0.099$&$0.052$&$0.011$\tabularnewline
LRT-EXT&$0.072$&$0.038$&$0.008$&&$0.094$&$0.052$&$0.012$&&$0.104$&$0.060$&$0.012$\tabularnewline
LRT-PBS[-1,1] & $0.086$& $0.040$& $0.008$&& $0.097$& $0.057$ &$0.011$ &&$0.070$& $0.040$ & $0.008$\tabularnewline
LRT-PBS[-2,2] & $0.098$& $0.048$& $0.012$&& $0.102$& $0.046$ &$0.008$ &&$0.106$& $0.056$ & $0.013$
\tabularnewline
\hline
\end{tabular}
\end{center}
\caption{Size Performance for Various Tests for Homogeneity of the Gaussian Location Parameter: Independent samples of different sizes are generated from $\mathcal{N}(1,1)$. We consider test for homogeneity versus general alternative. The EM test is as proposed in \protect\citeasnoun{ChenLi} using the R code provided on the second author's webpage \protect \url{http://sas.uwaterloo.ca/~p4li/software/index.html} of the EM test for Gaussian mixture with known variance. The $C(\alpha)$ test uses critical values from $\frac{1}{2} \chi_{0}^{2} + \frac{1}{2} \chi_{1}^{2}$ null  distribution. LRT-EXT uses bootstrap critical values tabulated in Table \ref{tab:bcval}. Results are based on 6,000 repetition. LRT-PBS (with stated support used for estimating the general mixture model) uses parametric bootstrap critical values with 500 bootstrap repetitions on the pre-specified support for the location parameter. }
\label{tab:sizenorm}
\end{table}

\section{Neyman $C(\alpha)$ Tests for Mixture Models} 
Neyman's $C(\alpha)$ tests can be viewed as an expanded class of Rao (score) 
tests that accommodate general methods of estimation for nuisance parameters.  
In regular likelihood settings $C(\alpha)$ tests are constructed 
from the usual score components which consist of the first order logarithmic derivative of the likelihood. 
The $C(\alpha)$ tests can be shown to be asymptotically locally optimal and the associated 
regularity conditions for these results were originally given by 
\citeasnoun{Neyman59} and extended by \citeasnoun{BuhlerPuri} 
employing variants of the classical \Cramer conditions. In applying the $C(\alpha)$ approach to test 
for homogeneity in mixture models, the test statistics typically still take a simple form although 
their theory requires some substantial amendment due to the singularity of the score function. 
\citeasnoun{Gu} shows that the locally asymptotic normal  (LAN) apparatus of LeCam 
can be brought to bear to establish the large sample behavior and asymptotic optimality of 
the $C(\alpha)$ test for homogeneity.  The LeCam approach has two salient advantages:  it avoids making
superfluous further differentiability assumptions on the density, and it removes any need for
the symmetry assumption on the distribution of the heterogeneity that frequently
appears in earlier examples of such tests.  See e.g. \citeasnoun{Moran1973} and \citeasnoun{Chesher1984}.

The following two examples illustrate the construction of the $C(\alpha)$ test for parameter homogeneity in the Gaussian mixture model and the Poisson mixture model. Both tests lead to an over-dispersion test. In the Gaussian case, the test compares the sample variance with the variance under the null hypothesis. In the Poisson case, we reject the null of homogeneity if there exists over-dispersion in the sample variance in comparison to the sample mean.

\begin{example}\label{eg: gaussian}
  Consider testing for homogeneity in the Gaussian location mixture model with
  independent observations $X_i \sim \NN(\mu_i , 1), i = 1, \cdots , n$.
  Assume that $\mu_{i} = \mu_{0} + \tau \xi U_{i}$, for known $\tau$,
  and iid $U_i \sim F$ with $\EE U = 0$ and $\Var(U) = 1$.  The heterogeneity in $\mu_i$ is introduced via the random variable $U$. We would like to test homogeneity of $\mu_i$, 
  $H_0: \xi = 0$, with the location parameter $\mu_0$ treated as a nuisance
  parameter.  As mentioned earlier, the first-order logarithmic derivative for $\xi$ is degenerately zero, however we can construct the test statistics using its second-order derivative, which is found to be, 
$\nabla_{\xi}^{2} \log p(x| \mu_{0}, \xi=0)
=  \tau^2 ((x-\mu_{0})^{2} -1)$. 
The first-order score for the nuisance parameter $\mu_{0}$ is, 
$ \nabla_{\mu_{0}} \log p(x| \mu_{0}, \xi=0) = (x-\mu_{0}).$
Note that under the null, 
$\mathop{\rm cov} (\nabla_{\xi}^{2} \log p(X| 0, \mu_{0}), 
\nabla_{\mu_{0}} \log p(X| 0, \mu_{0}))=0$, thus the $C(\alpha)$ test statistics require no modification of the test statistics to reflect the fact that we need to estimate the nuisance parameter $\mu_0$ and 
thus, we have the locally asymptotically optimal $C(\alpha)$ test as
\[
Z_{n}=\frac{1}{\sqrt{2n}} \sum_{i=1}^n ((X_{i}-\mu_{0})^{2} - 1)
\]
The obvious estimate for the nuisance parameter is 
the sample mean, and we reject the null hypothesis when 
$(0 \vee Z_{n})^{2} > c_\alpha$ where $c_\alpha$ is the $(1-\alpha)$ quantile of $\frac{1}{2}\chi_{0}^{2} + \frac{1}{2}\chi_{1}^{2}$. The test statistic $Z_{n}$ depends on the sample variance of $X$. Under the general alternative model, we have $\Var(X) = \E_{\mu}[\Var(X|\mu)] + \Var_{\mu}[\E(X|\mu)] = 1 + \Var(\mu)$. Under the alternative, the magnitude of $Z_{n}$ solely depends on $\sqrt{n} \Var(\mu)$. 
\end{example}

\begin{example}\label{eg: pois}
  Consider now testing for homogeneity of the mean parameter in the Poisson model with
  independent observations $X_i \sim p(\cdot |\lambda_i), i = 1, \cdots , n$ with $p(x| \lambda) = \frac{\lambda^{x} \exp(-\lambda)}{x!}$.
  Assume that $\lambda_{i} = \lambda_{0}\exp( \tau \xi U_{i})$, for known $\tau$,
  and iid $U_i \sim F$ with $\EE U = 0$ and $\Var(U) = 1$.  We would like to test
  $H_0: \xi = 0$ with the mean parameter $\lambda_0$ treated as a nuisance
  parameter.  The second-order score for $\xi$ is found to be, 
$\nabla_{\xi}^{2} \log p(x| \lambda_{0}, \xi=0)
=  \tau^2 ((x-\lambda_{0})^{2} -\lambda_0)$
and the first-order score for $\lambda_{0}$ is, 
$ \nabla_{\lambda_{0}} \log p(x| \lambda_{0}, \xi=0) = (x-\lambda_{0})/\lambda_0.$
Note that under the null, 
$\mathop{\rm cov} (\nabla_{\xi}^{2} \log p(X| \lambda_{0},0), 
\nabla_{\lambda_{0}} \log p(X| \lambda_{0},0))=\lambda_0$.
Thus, we have the locally asymptotically optimal $C(\alpha)$ test as
\[
Z_{n}=\frac{1}{\sqrt{2n}} \sum_{i=1}^n \frac{((X_{i}-\lambda_{0})^{2} - \lambda_0 - (X_{i} - \lambda_0))}{\lambda_0}
\]
The obvious estimate for the nuisance parameter $\lambda_0$ is 
the sample mean $\bar X$, which further reduces $Z_{n} =\frac{1}{\sqrt{2n}} \sum_{i=1}^n\frac{((X_{i}-\bar X)^{2} - \bar X)}{\bar X}$  and we reject the null hypothesis when 
$(0 \vee Z_{n})^{2} > c_\alpha$. The test statistic $Z_{n}$ depends on the ratio of the sample variance and sample mean of $X$. Under the alternative model, we have $\Var(X) = \E(\lambda) + \Var(\lambda)$ and $\E(X) = \E(\lambda)$. The magnitude of the test statistics $Z_{n}$ under the alternative is determined by the ratio $\sqrt{n} \Var(\lambda)/\E(\lambda)$.
\end{example}

\section{The EM Test of Homogeneity for Finite Mixture Models}
The $C(\alpha)$ test described above is very attractive because its test statistic is easy to construct 
under the null model and its asymptotic theory is also relatively simple. The recently proposed EM test 
of \citeasnoun{ChenLi}, \citeasnoun{CLM} and \citeasnoun{LiChen} shares these nice features. 
The EM test employs a penalized log likelihood ratio statistic, and instead of optimizing over general
class of heterogeneous alternatives optimization is restricted to a smaller finite dimensional class.
Given the mixture model \eqref{Mdens}, we consider finite mixing distributions
$\eta = \sum_{h=1}^{m} \alpha_{h} \delta(\mu_h)$ with $m$ distinct support points at locations 
$\{\mu_1, \dots, \mu_m\}$. We are interested in testing $H_0: m = 1 \mbox{\ \ versus\ \ } H_{A}: m > 1$.
Rather than consider the full panoply of alternatives, attention is restricted to mixing distributions
with only two points of support,
\[
\Omega_{2}( \beta) = \{   \beta \delta(\mu_1) +  (1- \beta) \delta(\mu_2): \mu_{1}, \mu_{2} \in I \}
\]
the relative mass of the two support points, $\beta \in (0, 0.5]$, is bounded away from zero by the 
penalized log likelihood,
\[
pl_n(\Psi) = \sum_{i=1}^{n} \log p_\Psi(X_i) + P(\beta)
\]
where $\Psi \in \Omega_{2}(\beta)$, and $P(u)  = C \log(1 - | 1 - 2 u | )$.  The set $I$ over which
the $\mu$'s are optimized is taken to be the support of the observations in the Gaussian location
mixture setting.  Optimization is carried out via the EM algorithm over the three parameters,
$\{ \beta, \mu_1, \mu_2 \}$, and the test statistic is,
\[
M_{n} = 2 \{ pl_{n}(\hat \Psi) - \sum_{i} \log p_{\tilde \Psi}(X_i)\}, 
\]
where $\hat \Psi$ and $\tilde \Psi$  denote estimates for the model under the alternative and null, respectively. 
Selection of tuning parameters including initial values and stopping criteria for the EM procedure may, 
of course, influence performance.
Penalization has the desirable effect of avoiding the singularity that would otherwise occur as
$\beta \rightarrow 0$.  $M_n$ has been shown to have a $\frac{1}{2}\chi_{0}^{2} + \frac{1}{2} \chi_{1}^{2}$
limiting distribution.  Testing for additional mixture components yields more complicated mixtures of $\chi^2$'s.
In the next section we compare the size and power performance of our general LRT with the EM test and the 
$C(\alpha)$ test for different mixture models in simulations.


\section{Some Simulation Evidence}\label{sec:sim}
To compare power of the $C(\alpha)$, the EM test and LRT to detect heterogeneity in the Gaussian location
model we conducted five distinct experiments.  Two were based on variants of the 
\citeasnoun{Chen.95} example with the discrete  mixing distribution 
$\eta = (1-\lambda) \delta(a + h/(1-\lambda)) + \lambda \delta(a -h/\lambda)$.  In the
first experiment we set $\lambda = 1/3$, as in the original Chen example, in the second
experiment we set $\lambda = 1/20$ and in both experiments, $a$ is set to be zero. The sample size is fixed at $n = 200$. We consider
five tests  
\begin{enumerate}
\item[(i)] the $C(\alpha)$ as described in Example \ref{eg: gaussian}. Under $H_0: h=0$, the nuisance parameter $a$ can be estimated by the sample mean. 
\item[(ii)] a parametric version of the 
LRT in which only the values of $a$ and $h$ are assumed to be unknown and the relative probabilities 
associated with the two mass points are known;  this enables us to relatively easily find 
the MLE: profiling out $a$ first, $\hat h$ can be estimated by separately optimizing the likelihood on the positive and negative 
half-line and taking the best of the two solutions; and then we can find the best pair of $(\hat a, \hat h)$ that maximizes the likelihood. 
\item[(iii)] the Kiefer-Wolfowitz LRT computed with equally spaced binning of 300 grid points on the support of the sample
\item[(iv)]  the classical Kolmogorov-Smirnov test of normality 
\item[(v)]  the EM test for one component versus two components. 
\end{enumerate}
All of the power comparisons are based on 10,000 simulation replications. We consider 21 distinct values of $h$ for each of the experiments equally spaced on the respective
plotting regions.  

\begin{figure}[h]
  \centering
  \subfloat{\includegraphics[width=.45\textwidth]{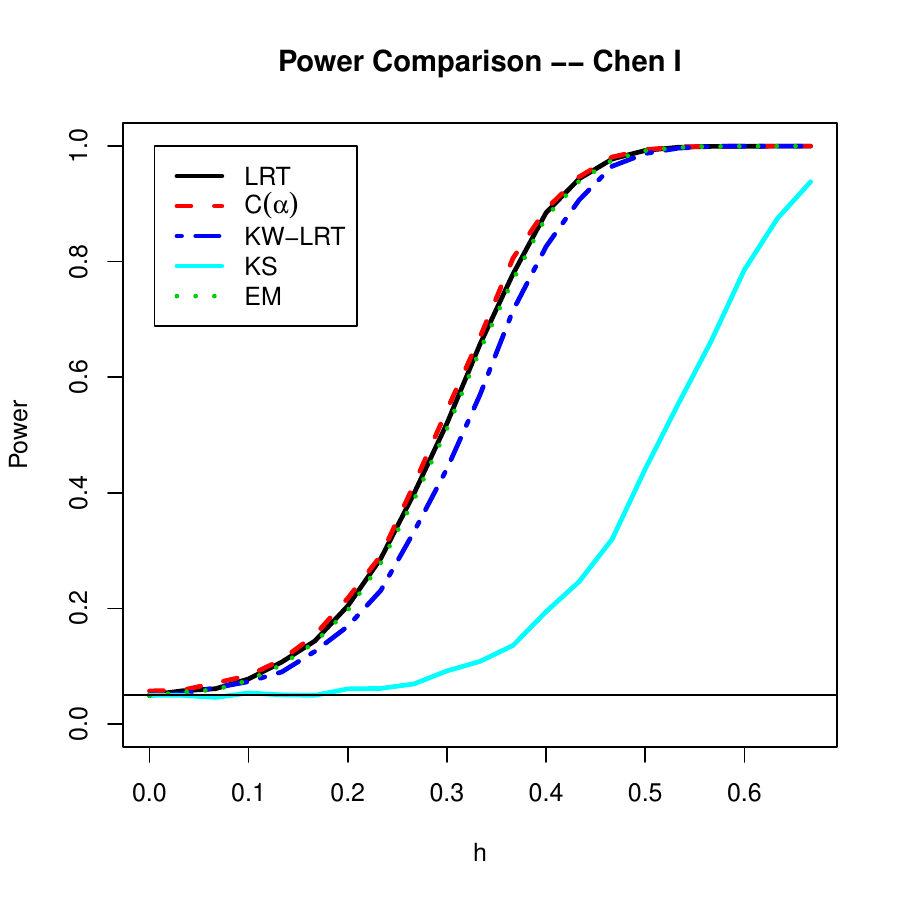}} \qquad
  \subfloat{\includegraphics[width=.45\textwidth]{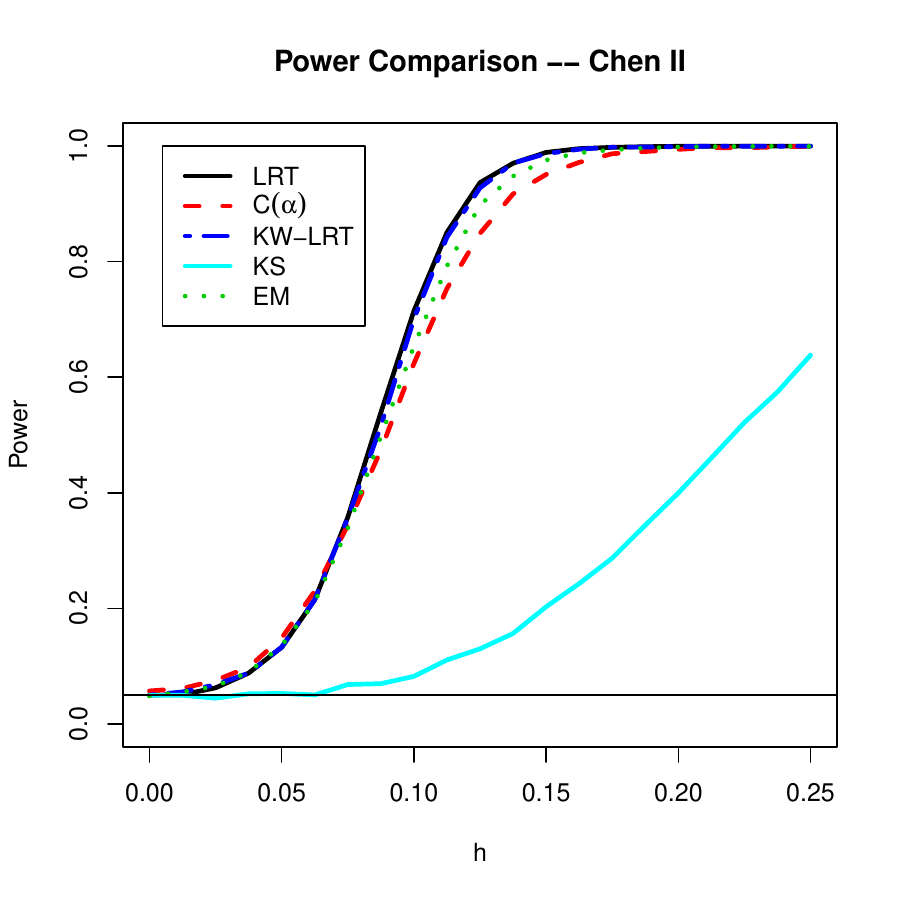}}
\caption{Power Comparison of Several Tests of Parameter Homogeneity:
The left panel illustrates empirical power curves for four tests of parameter
homogeneity for the Chen (1995) mixture with \ensuremath{\lambda = 1/3}, in the right
panel we illustrate the power curves for the same four tests for the Chen mixture with 
\ensuremath{\lambda = 1/20}.
Note that in the more extreme (right) setting, the LRTs outperform 
the \ensuremath{C(\alpha)} test.}
\label{ChenFigs}
\end{figure}

In the left panel of Figure \ref{ChenFigs} we illustrate the results for the first experiment with $\lambda = 1/3$:  With the location invariance property of the Gaussian mixture model, we use the bootstrap critical values in Table \ref{tab:bcval} for the nonparametric LRT.  The EM test, $C(\alpha)$ and the parametric LRT are essentially indistinguishable in
this experiment,  and each has slightly better performance than the nonparametric LRT.
All four of these tests perform substantially better
than the Kolmogorov-Smirnov test.  In the right panel of Figure \ref{ChenFigs} we have results of another
version of the Chen example, except that now $\lambda = 1/20$, so the mixing distribution
is much more skewed.  Still $C(\alpha)$ does well for small values of $h$, but for $h \geq 0.07$
the two LRT procedures, which are now essentially indistinguishable, dominate. The performance of the EM test lies in between the $C(\alpha)$ test and the nonparametric LRT test. 
Again, the KS test performance is poor compared to the other tests explicitly designed 
for the mixture setting. 

\begin{figure}[h]
  \centering
  \subfloat{\includegraphics[width=.45\textwidth]{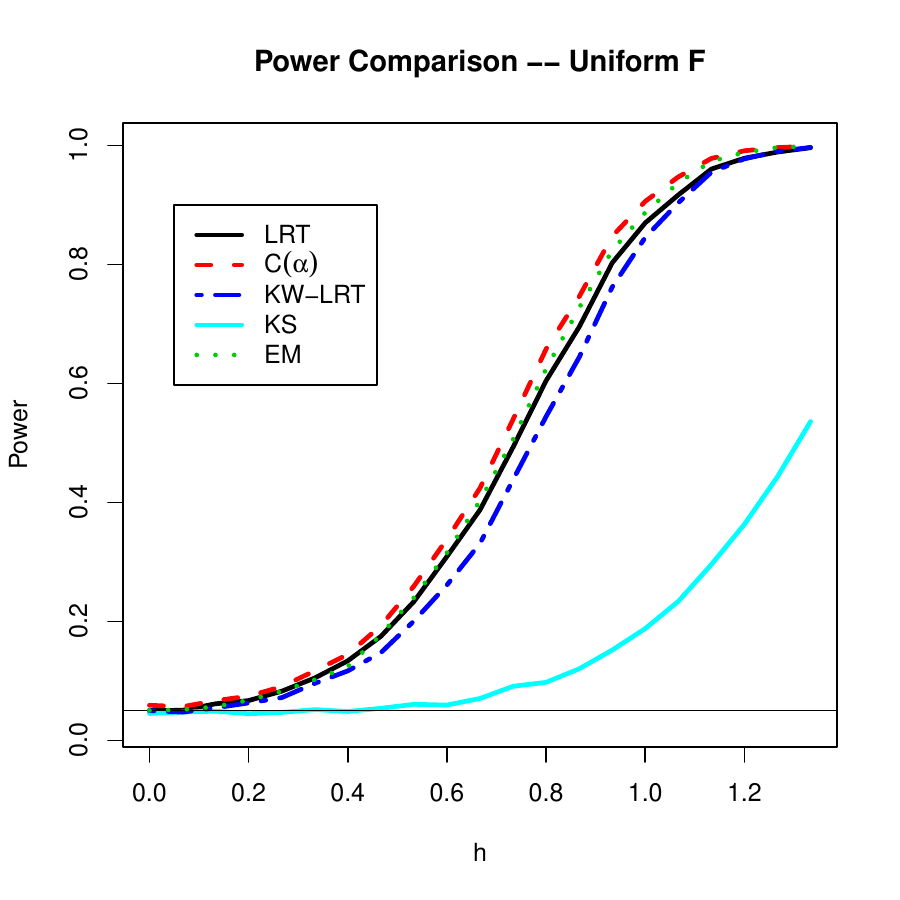}} \qquad
  \subfloat{\includegraphics[width=.45\textwidth]{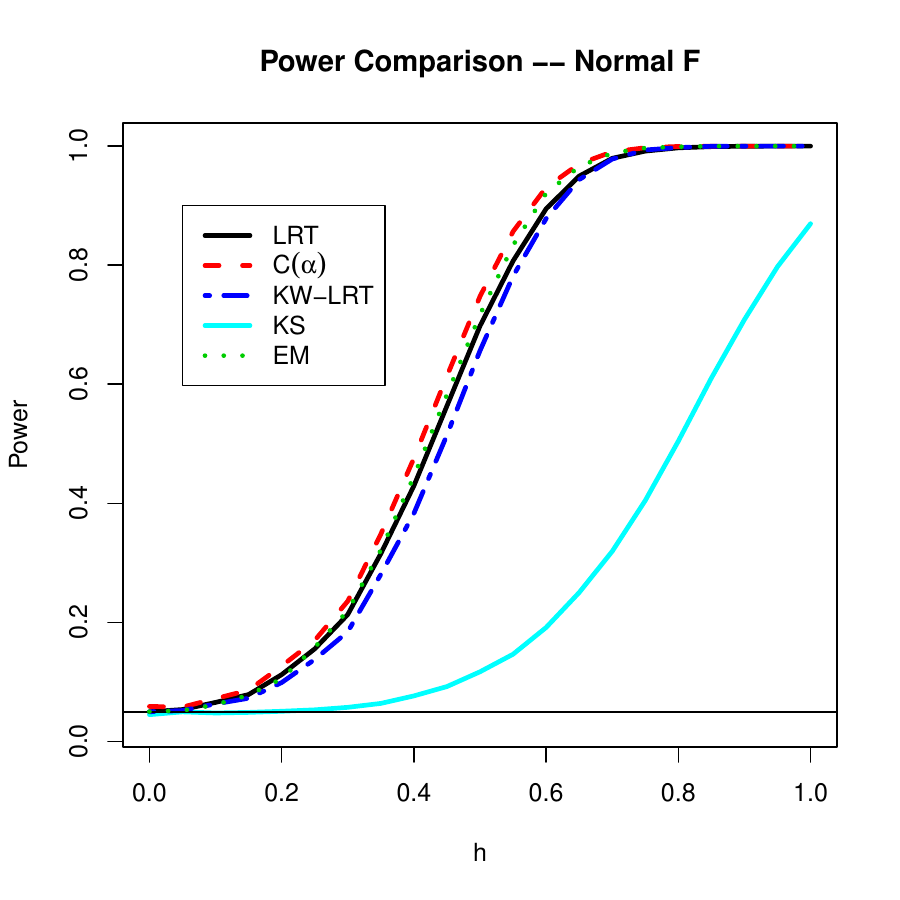}}
 \caption{Power Comparison of Several Tests of Parameter Homogeneity:
 The left panel illustrates empirical power curves for four tests of parameter
 homogeneity for uniform mixtures of Gaussians 
 with \ensuremath{\vartheta} on \ensuremath{[-h,h]}, on the
 right panel the same four power curves are depicted for Gaussian mixtures of
 Gaussians with standard deviation \ensuremath{h}.}
\label{UGFigs}
\end{figure}

In Figure \ref{UGFigs} we illustrate the results of two additional experiments, both of which are
based on smooth mixing distributions with densities with respect to Lebesgue measure and a sample size of $n = 200$. On the left we consider the uniform distribution on the interval $[-h,h]$. Here we can reduce the parametric LRT to optimizing over the positive half-line to compute the MLE, $\hat h$. This would seem to give the parametric
LRT a substantial advantage over the Kiefer-Wolfowitz nonparametric MLE, however as is
clear from the figure there is little difference in their performance.  Again, the $C(\alpha)$
test and the EM test are somewhat better than either of the LRTs, but the difference is modest.
In the right panel of Figure \ref{UGFigs} we have a similar setup, except that now the mixing distribution
is Gaussian with scale parameter $h$, and again the ordering is very similar to the uniform
mixing case. In all of these experiments, since the asymptotic behavior of the parametric LRT is unknown, we  use its empirical critical values under the null. 

In the last simulation experiment on testing for homogeneity in a normal model we consider data that are generated from a two-component mixture of the form
\[
(1-\alpha) N(\theta_1,1) + \alpha N(\theta_2, 1)
\]
with a very small value of $\alpha$. This is the second local alternative model considered by \citeasnoun{Chen16}. Notably, this also fits the discussion of local alternative model on page 94 in \citeasnoun{Lindsay95}. In simulation, we fix $\alpha = 0.005$, $\theta_1 =\theta_0= 0$ and $\theta_2 = b$ and conduct two sets of experiments. The first fixes $\theta_2 = -4.5$ and allows the sample size $n$ to change and the second varies values of $\theta_2$ for fixed sample size $n = 400$. Results are reported in Table \ref{tab: local}. We find that in all settings, the LRT outperforms both $C(\alpha)$ and the EM test by a considerable margin, with the EM test having advantages compared to $C(\alpha)$. This suggests that for detecting small mass points away from the main bulk of the data the LRT is the method of choice. This kind of behavior is also observed in the empirical example in Section~\ref{sec:eg}, where only the LRT is able to detect deviations from homogeneity.

A theoretical explanation for the findings in this experiment can be obtained by considering the likelihood expansion corresponding to a specific type of local alternative. Adopting the notation in \citeasnoun{Chen16} let $\alpha := \eta/\sqrt{n}$, $\theta_{1n} := \theta_0 - n^{-1/2} \tau (\frac{\eta}{1 - n^{-1/2}\eta})^{1/2}$ and $\theta_{2n} := \theta_0 + \tau (\frac{1-n^{-1/2} \eta}{\eta})^{1/2} \to \theta_0 + \tau /\sqrt{\eta} \equiv \theta_2$. As shown in \citeasnoun{Chen16} the likelihood ratio expansion in this case takes the form 
\[
\frac{\eta}{\sqrt{n}} \sum_{i} W_i - \frac{1}{2} \frac{\eta^2}{n} \sum_{i} W_i^2 + o_P(1) 
\]
with 
\[
W_i = \frac{f(x_i, \theta_2)-f(x_i, \theta_0)}{f(x_i, \theta_0)} - \frac{\tau}{\sqrt{\eta}} \frac{f'(x_i, \theta_0)}{f(x_i,\theta_0)}
\] 
provided $W_i$ is square integrable. Note that $W_i \approx \frac{\tau^2}{2\eta} f''(X_i, \theta_0)/f(X_i, \theta_0)$ only if $\theta_2$ is very close to $\theta_0$. This already suggests that the asymptotic optimality of the $C(\alpha)$ for detecting local alternatives will only continue to hold for $\tau \approx 0$. This helps to explain the clear advantages of we observe for LRT and EM tests when compared to the performance of $C(\alpha)$ in these extreme cases. 



\begin{figure}[h]
  \centering
  \subfloat{\includegraphics[width=.45\textwidth]{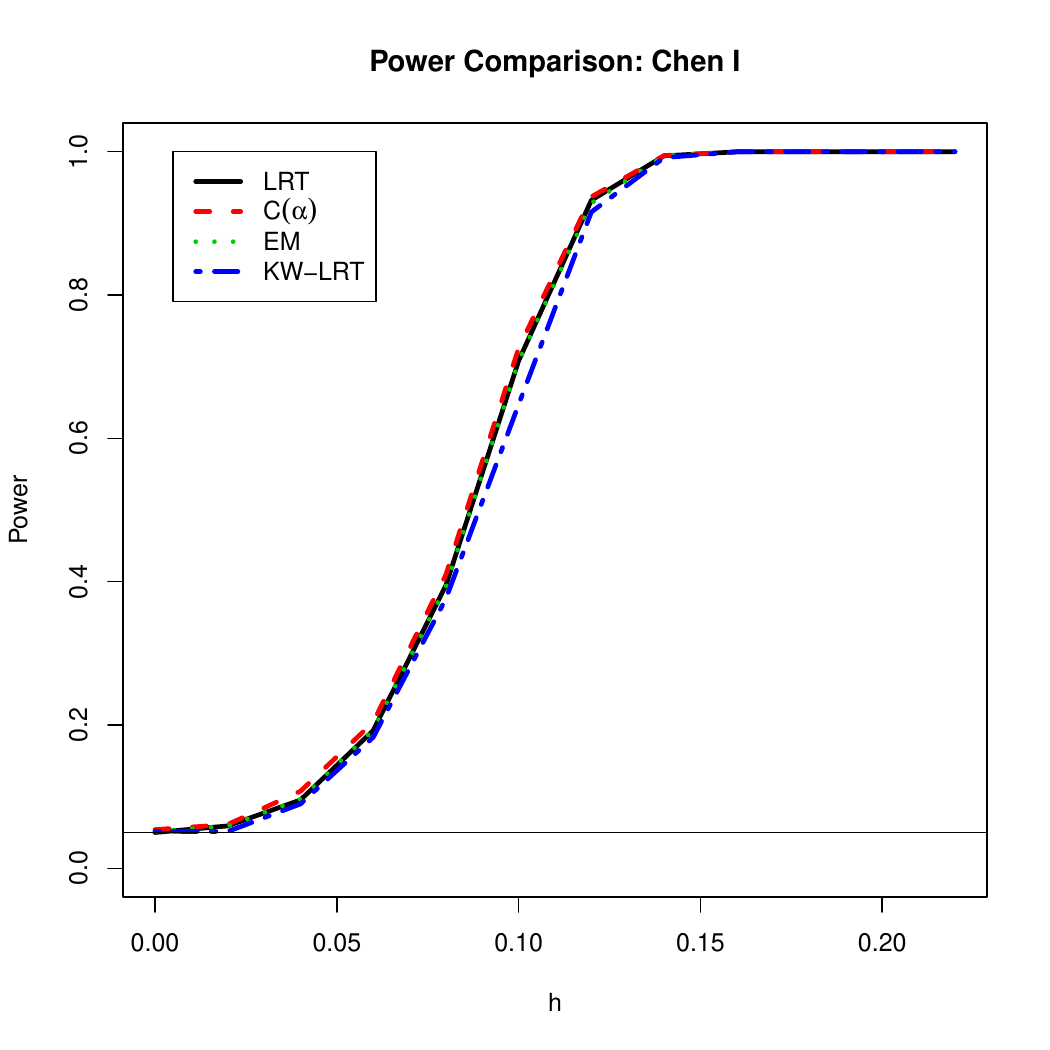}} \qquad
  \subfloat{\includegraphics[width=.45\textwidth]{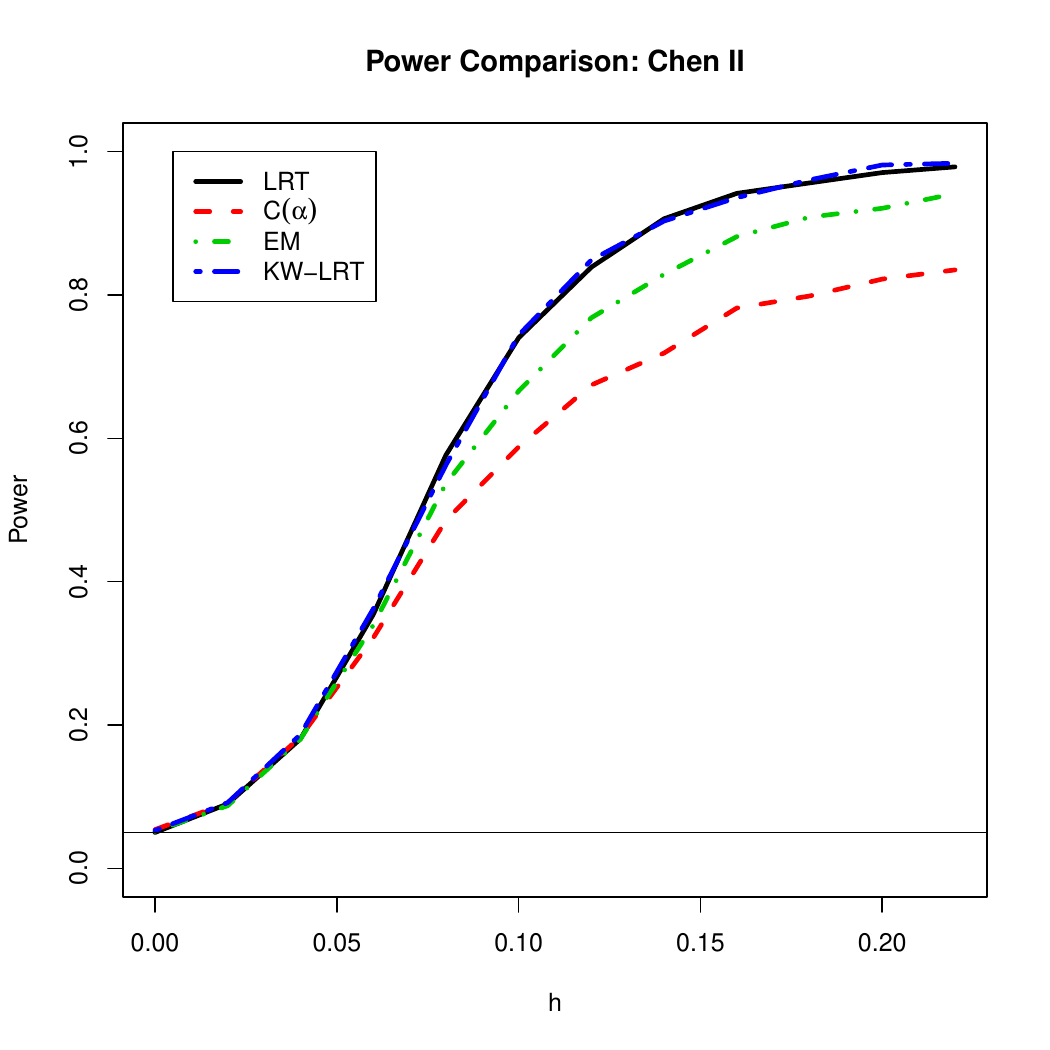}}
 \caption{Power Comparison of Several Tests of Parameter Homogeneity for Poisson Mixture Models:
 The figure illustrates empirical power curves for three tests of parameter
 homogeneity for a discrete mixtures of Poisson. The discrete mixing distribution is specified as $F(\mu) = (1-\lambda) \delta(2\exp(h/(1-\lambda))) + \lambda \delta(2\exp(-h/\lambda))$ with $\lambda = 1/3$ in the left panel and $\lambda=1/20$ in the right panel for $h$ taking 21 different values. The critical values for LRT are based on the bootstrap method. The empirical power curve is based on 5,000 repetitions.}
\label{PoisFigs}
\end{figure}

We also consider the power performance of the the above mentioned tests for Poisson mixture models except for the Kolmogorov-Smirnov test. Similarly to the Gaussian case, the Poisson mean parameter has the discrete mixing distribution 
$\eta = (1-\lambda) \delta(a\exp(h/(1-\lambda))) + \lambda\delta(a\exp((-h/\lambda))$. We consider $\lambda = 1/3$ and $\lambda=1/20$ case and set $a = 2$ in both cases. The $C(\alpha)$ test is constructed as described in Example \ref{eg: pois} with $H_0: h=0$ and $a$ as the nuisance parameter. Since the Poisson distribution does not take a location shift form, we resort to the parametric bootstrap method described in Section \ref{sec: bcval} to determine the critical value with a bounded support on $(0, 4)$ for the mean parameter with 5,000 repetition. To speed up simulation, we also adopt the warp bootstrap method in \protect \citeasnoun{warp}. Figure \ref{PoisFigs} shows the power for the $C(\alpha)$ test, the EM test and the KW-LRT for different values of $h$. Again, we observe similar pattern of the power curves as in the Gaussian case. For more extreme mixing distribution, the KW-LRT dominates the other two tests by quite a substantial margin. 

\begin{figure}[h]
  \centering
  \subfloat{\includegraphics[width=.45\textwidth]{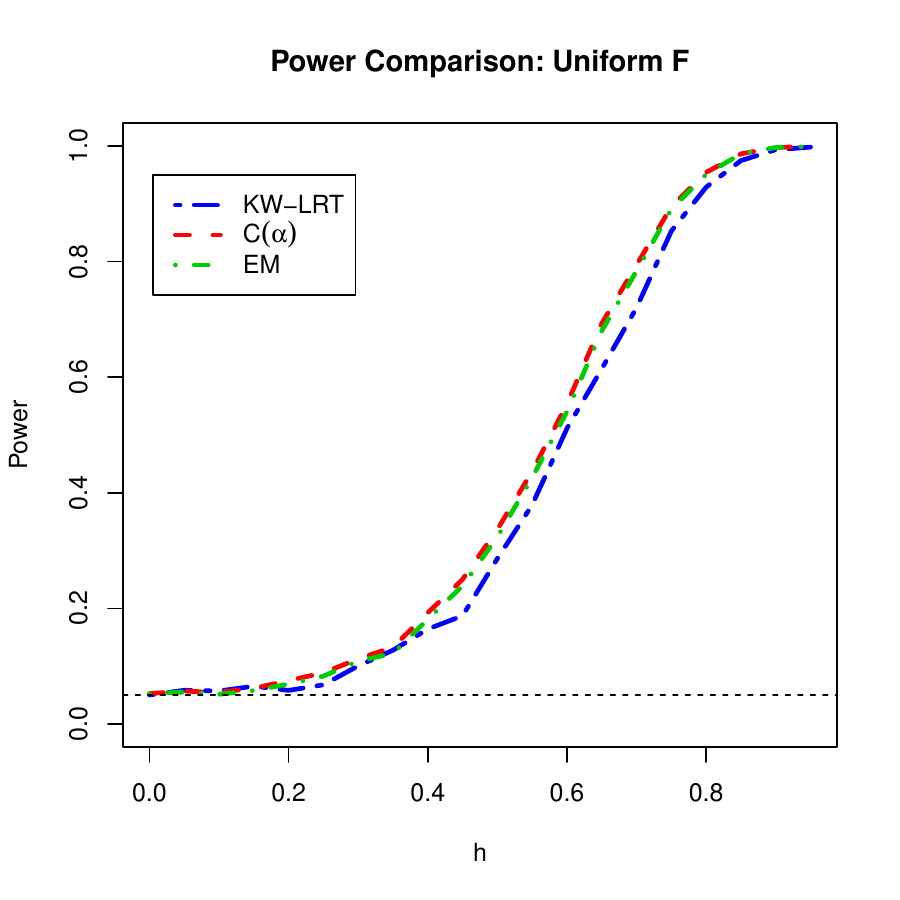}} \qquad
  \subfloat{\includegraphics[width=.45\textwidth]{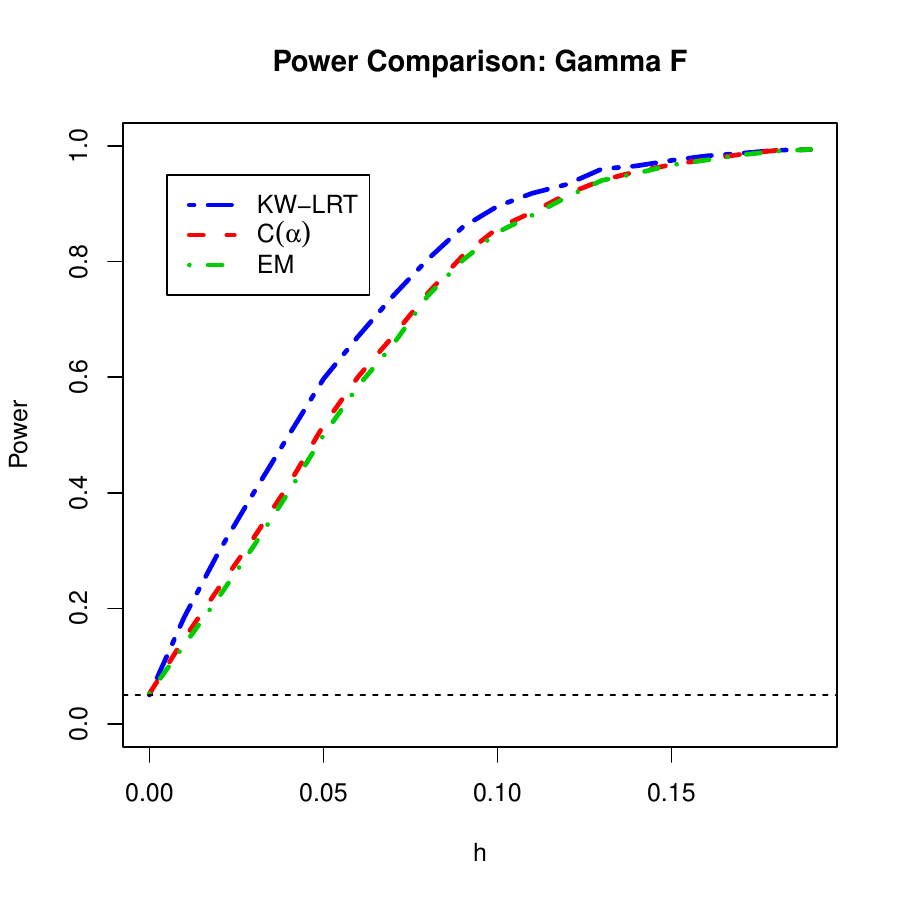}}
 \caption{Power Comparison of Several Tests of Parameter Homogeneity for Poisson Mixture Models:
The left panel illustrates empirical power curves for three tests of parameter
 homogeneity for uniform mixtures of Poissons 
 with $\lambda = 2\exp(k)$ and $k$ follows uniform distribution on \ensuremath{[0,h]}, on the
 right panel the same three power curves are depicted for Gamma mixtures of
 Poissons with $\lambda = 2\exp(k)$ and $k$ follows Gamma distribution with shape parameter $h$ and scale parameter $1/2$. Results are based on $n = 1,000$ and 5,000 simulation repetition.}
\label{CPFigs}
\end{figure}

In Figure \ref{CPFigs} we illustrate the results for Poisson mixtures with continuous mixing distribution. In both experiments, the mean parameter is set to be $2 \exp(k)$ where $k$ has a continuous distribution. On the left, we consider $k$ following a uniform distribution on $[0, h]$ with $h$ taking 20 distinct equally spaced values on $[0, 0.95]$. On the right, we have $k$ following a Gamma distribution with shape parameter $h$ and scale parameter 1/2 and $h$ taking 20 distinct equally spaced values on $[0, 0.19]$. The KW-LRT performs slightly worse than $C(\alpha)$ and the EM tests for the uniform case, but dominates the other two for the Gamma case. 

\begin{table}[!tbp]
\begin{center}
\begin{tabular}{lrrrrrrr}
\hline\hline
\multicolumn{1}{l}{}&\multicolumn{1}{c}{n = 200}&\multicolumn{1}{c}{n = 400}&\multicolumn{1}{c}{n = 800}&\multicolumn{1}{c}{b = -6}&\multicolumn{1}{c}{b= -4}&\multicolumn{1}{c}{b = -2}&\multicolumn{1}{c}{b = -1}\tabularnewline
\hline
LRT&$0.536$&$0.770$&$0.935$&$0.866$&$0.680$&$0.128$&$0.061$\tabularnewline
EM&$0.354$&$0.508$&$0.715$&$0.703$&$0.412$&$0.090$&$0.054$\tabularnewline
$C(\alpha)$&$0.296$&$0.412$&$0.578$&$0.635$&$0.329$&$0.093$&$0.060$\tabularnewline
\hline
\end{tabular}\end{center}
 \caption{Power Comparison of Several Tests of Parameter Homogeneity for two-component Normal Mixture Models: Results in column two to four are proportion of rejection of homogeneity using data generated from $0.995 \mathcal{N}(0, 1) + 0.005 \mathcal{N}(-4.5, 1)$ with various sample size $n$ stated as the column names. Results in column five to eight are proportion of rejection of homogeneity using a sample of size 400 generated from $0.995 \mathcal{N}(0,1) + 0.005 \mathcal{N}(b, 1)$ with $b$ taking different values stated as the column names. The empirical power is based on 10,000 repetitions and LRT uses tabulated critical values of  $5\%$ nominal size.} 
 \label{tab: local}
\end{table}

\section{Empirical Example} \label{sec:eg}
\begin{figure}[h]
  \centering
\includegraphics[scale = 0.5]{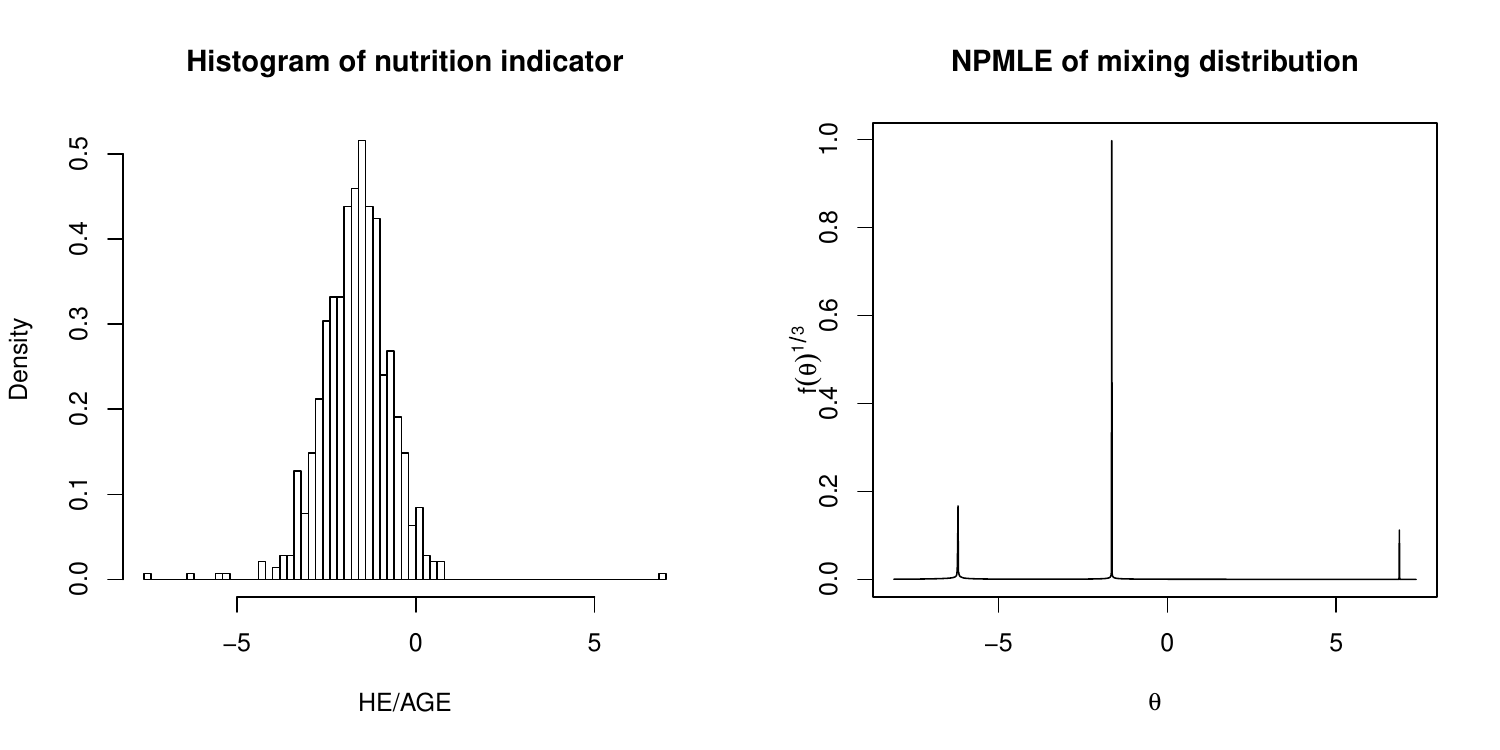}
\caption{Thai Preschool Children Nutritional Status: The left panel plots the histogram of the HE/AGE data of size 708. The right panel depicts the Kiefer-Wolfowitz nonparametric maximum likelihood estimator of the mixing distribution for the location parameter of the normal mixture model with 1500 grid points. The cube root of the mass associated with the support points are plotted in an effort to render the small masses more visible.}
\label{fig: example}
\end{figure}

We briefly revisit an application considered in \citeasnoun{Bohning92} and \citeasnoun{Chen16} on modeling a nutritional indicator in order to detect subclinical malnourishment. To evaluate nutritional status of children in developing countries, a standardized height score (HE/AGE) is often used. It is defined as height of the child recentered by the median and normalized by the standard deviation of heights for a reference population of the same age and sex. Under the hypothesis of no malnutrition, we expect the data to follow a normal distribution with unit variance. Deviation from homogeneous normal distribution provides evidence for malnutrition of the group of children. We conduct nonparametric LRT, EM test and the $C(\alpha)$ test for homogeneity of the location parameter. Both the EM and the $C(\alpha)$ test find insufficient evidence against homogeneity, with EM test reporting a p-value close to 1 and the $C(\alpha)$ test statistic taking a value 0. In contrast, the nonparametric LRT finds strong evidence against homogeneity.  Adopting the parametric bootstrap method and restricting the support to between the  5-th and 95-th percentile of the data, the nonparametric likelihood ratio test statistic equals 12.77, while the parametric bootstrap critical value at $5\%$ level equals 4.68. The nonparametric LRT using an unrestricted support and tabulated critical values leads to the same conclusion. 
Figure \ref{fig: example} shows the histogram of the data and the nonparametric MLE for the mixing distribution of the location parameter based on estimation method described in Section \ref{sec:KWE}. The vast majority of the mass (0.993) is allocated to the point -1.64 but we find two additional mass points at -6.19 and 6.87 with associated mass 0.005 and 0.002. Clearly, the largest data point has a mass of its own, while the mass point at -6.19 captures the very small proportion of observations at the left tail of the histogram. Although both mass points are small, they provide overwhelming evidence against homogeneity which is surprisingly not picked up by either EM or $C(\alpha)$ test. This sheds new light into the nature of our competing tests and illustrates that the LRT is particularly well suited to detecting deviations from the null which correspond to small mass points at extreme locations  lending further support to our simulation results.

\section{Conclusion}
We have seen that the Neyman $C(\alpha)$ test provides a simple,  powerful, albeit somewhat
irregular, strategy for constructing tests of parameter homogeneity. In contrast, the development of likelihood ratio testing for mixture models has been
somewhat inhibited by their apparent  computational difficulty, as well as the complexity
of their asymptotic theory.  Recent developments in convex optimization have dramatically 
reduced the computational effort of earlier EM methods, and new theoretical developments
have led to practical simulation methods for large sample critical values for the 
Kiefer-Wolfowitz nonparametric version of the LRT.  Local asymptotic optimality of
the $C(\alpha)$ test assures that it is highly competitive in many circumstances, but
we have illustrated a class of examples where the LRT has a slight edge.  The EM tests
of \citeasnoun{LiChen} provide an intermediate approach relying on a more restricted
formulation of the likelihood.  The approaches
are complementary;  clearly there is little point in testing for heterogeneity if there is
no mechanism for estimating models under the alternative.  Our LRT approach obviously
provides a direct pathway to estimation of the mixture model under general alternatives. 
Since parametric mixture models are notoriously tricky to estimate, it is a remarkable fact that 
the nonparametric formulation of the MLE problem \`{a} la Kiefer-Wolfowitz can be solved quite efficiently --
even for large sample sizes by binning -- and effectively used as an alternative testing
procedure.  We hope that these new developments will encourage others to explore these methods.

\bibliography{lrt}

\begin{appendix}
\section{Technical details}

\textbf{Proof of (\ref{approx})} 
Given a measure $\eta \in \Gc, \eta \neq \delta(0)$ define $V(\eta) := \sum_{k=2}^\infty \frac{\kappa_k^2(\eta)}{k!}$. Also, define for $n\in\N$ and $\alpha \in \R$ the probability measure $\tilde \eta_n := p_n\delta_{c_n} + (1-p_n)\eta$ with $p_n := 1 - V(\eta)/n$ and $c_n := \frac{1-p_n}{p_n}(\alpha - \kappa_1(\eta))$ [the dependence of $p_n,c_n$ on $\eta$ is suppressed in the notation]. Note that for any $N > 0$ there exists $n_0 \in\N$ such that for $n\geq n_0$ we have $\tilde \eta_n \in \Gc$ for all $\alpha \in [-N,N]$. Moreover, by construction $\kappa_1(\tilde \eta_n) = \alpha(1-p_n)$ and 
\[
\kappa_k(\tilde \eta_n) = \kappa_k(\eta)(1-p_n) + (1-p_n)\Big(\frac{1-p_n}{p_n}\Big)^{k-1}(\alpha-\kappa_1(\eta))^k
\] 
for $n \in \N$. This implies for $n \geq n_0$ with some $n_0$ independent of $\eta$ we have a.s.
\begin{align*}
\Big| \alpha Y_1 + \sum_{k=2}^\infty \frac{Y_k \kappa_k(\eta)}{(k!)^{1/2}} - \frac{1}{1-p_n}\sum_{k=1}^\infty \frac{Y_k \kappa_k(\tilde \eta_n)}{(k!)^{1/2}}\Big|
\leq & \frac{1-p_n}{p_n}\sum_{k=2}^\infty \frac{|Y_k|\tilde C^k}{\sqrt{k!}}\Big( \frac{1-p_n}{p_n}\Big)^{k-2}
\\
\leq & \frac{2\tilde C^{2} V(\eta)}{n} \sum_{k=2}^\infty \frac{|Y_k|}{\sqrt{k!}}
\end{align*}
and
\[
\Big| \alpha^2 + \sum_{k=2}^\infty \frac{\kappa_k^2(\eta)}{k!} -\frac{1}{(1-p_n)^{2}} \sum_{k=1}^\infty \frac{\kappa_k^2(\tilde \eta_n)}{k!}\Big|
\leq \frac{CV(\eta)}{n} 
\]
for finite constants $C,\tilde C$ depending only on $N$ but not on $\alpha$ and $\eta$ [note that $\eta \in \Gc$ has support contained in $[L,U]$].
Thus for every $N<\infty,\eps>0$ there exists $n_0$ independent of $\eta$ such that for all $n \geq n_0$ we have with probability at least $1-\eps$ 
\[
\sup_{\eta \in \Gc} \frac{\sum_{k=1}^\infty \frac{Y_k \kappa_k(\eta)}{(k!)^{1/2}}}{\Big(\sum_{k=1}^\infty \frac{\kappa_k^2(\eta)}{k!}\Big)^{1/2} }
 \geq \sup_{\alpha\in [-N,N]} \sup_{\eta \in \Gc}\frac{\alpha Y_1+\sum_{k=2}^\infty \frac{Y_k \kappa_k(\eta)}{(k!)^{1/2}} }{\Big(\alpha^2+\sum_{k=2}^\infty \frac{\kappa_k^2(\eta)}{k!}\Big)^{1/2}} - \eps. 
\]
Next, observe that for all $\eps >0$ there exists $N \in \R$ such that with probability at least $1-\eps$ 
\[
\sup_{\alpha\in \R\backslash [-N,N]} \sup_{\eta \in \Gc}\frac{\alpha Y_1+\sum_{k=2}^\infty \frac{Y_k \kappa_k(\eta)}{(k!)^{1/2}} }{\Big(\alpha^2+\sum_{k=2}^\infty \frac{\kappa_k^2(\eta)}{k!}\Big)^{1/2}} \leq |Y_1| + \eps.
\]
Finally, note that
\[
\sup_{\eta \in \Gc} \frac{\sum_{k=1}^\infty \frac{Y_k \kappa_k(\eta)}{(k!)^{1/2}}}{\Big(\sum_{k=1}^\infty \frac{\kappa_k^2(\eta)}{k!}\Big)^{1/2} } \geq |Y_1| \quad \mbox{a.s.}
\]
[consider the sequence of measures $\eta_n = \delta_{sign(Y_1)/n}\in \Gc$]. 

Summarizing the findings above, we have shown that for any $\eps>0$ we have with probability at least $1-2\eps$ 
\[
\sup_{\eta \in \Gc} \frac{\sum_{k=1}^\infty \frac{Y_k \kappa_k(\eta)}{(k!)^{1/2}}}{\Big(\sum_{k=1}^\infty \frac{\kappa_k^2(\eta)}{k!}\Big)^{1/2} }
\geq \sup_{\alpha\in \R}\sup_{\eta \in \Gc} \frac{\alpha Y_1+\sum_{k=2}^\infty \frac{Y_k \kappa_k(\eta)}{(k!)^{1/2}} }{\Big(\alpha^2+\sum_{k=2}^\infty \frac{\kappa_k^2(\eta)}{k!}\Big)^{1/2}} - \eps.
\]
By letting $\eps \to 0$ the above can be turned in an almost sure inequality with no $\eps$ on the right-hand side. Finally, 
setting $\alpha = \kappa_1(\eta)$ we see that the converse inequality also holds almost surely. Thus we have shown that
\[
\sup_{\eta \in \Gc} \frac{\sum_{k=1}^\infty \frac{Y_k \kappa_k(\eta)}{(k!)^{1/2}}}{\Big(\sum_{k=1}^\infty \frac{\kappa_k^2(\eta)}{k!}\Big)^{1/2} }
= \sup_{\alpha\in \R}\sup_{\eta \in \Gc} \frac{\alpha Y_1+\sum_{k=2}^\infty \frac{Y_k \kappa_k(\eta)}{(k!)^{1/2}} }{\Big(\alpha^2+\sum_{k=2}^\infty \frac{\kappa_k^2(\eta)}{k!}\Big)^{1/2}} \quad \mbox{a.s.}
\]
Define $\beta_k := \frac{\kappa_k(\eta)}{(k!)^{1/2}}$
and
\[
g_{Y,\eta}(\alpha) := \frac{\alpha Y_1+\sum_{k=2}^\infty \frac{Y_k \kappa_k(\eta)}{(k!)^{1/2}} }{\Big(\alpha^2+\sum_{k=2}^\infty \frac{\kappa_k^2(\eta)}{k!}\Big)^{1/2}}.
\] 
Fix a realization of $Y_1,Y_2,...$ and an $\eta \in \Gc$. Computing the derivative of $g_{Y,\eta}$ with respect to $\alpha$ shows that the function $g$ has a maximum at $
\alpha^* = Y_1\frac{\sum_{k=2}^\infty \beta_k^2}{\sum_{k=2}^\infty Y_k\beta_k},
$ if $\sum_{k=2}^\infty Y_k\beta_k > 0$ and that the supremum of $g_{Y,\eta}$ over $\alpha \in \R$ equals $Y_1^2$ if $\sum_{k=2}^\infty Y_k\beta_k \leq 0$. Some simple algebra shows that for $\sum_{k=2}^\infty Y_k\beta_k > 0$ we have
\[
g_{Y,\eta}(\alpha^*) = \Big(Y_1^2 + \frac{\Big(\sum_{k=2}^\infty Y_k\beta_k\Big)^2}{\sum_{k=2}^\infty \beta_k^2}\Big)^{1/2}.
\]
Thus we obtain
\[
\Big(\sup_{\eta \in \Gc} \Big(\frac{\sum_{k=1}^\infty \frac{Y_k \kappa_k(\eta)}{(k!)^{1/2}}}{\Big(\sum_{k=1}^\infty \frac{\kappa_k^2(\eta)}{k!}\Big)^{1/2} } \Big )_+\Big)^2
= Y_1^2 + \sup_{\eta \in \Gc} \frac{\Big( \Big(\sum_{k=2}^\infty \frac{Y_k \kappa_k(\eta)}{(k!)^{1/2}}\Big)_+ \Big)^2}{\sum_{k=2}^\infty \frac{\kappa_k^2(\eta)}{k!}}
\]
and this directly implies (\ref{approx}) \hfill $\Box$  \\

\textbf{Proof of Theorem \ref{th:iid}} The proof of the expansion in \eqref{eq:iid1} is very similar to the proof of \eqref{eq:main1} in Theorem 2.8, but much simpler since the data are i.i.d. and do not form a triangular array. For this reason we will only sketch the main arguments. First, observe that the class of functions $\Fc := \{s_{\eta,\mu_0}|\eta \in \Gc\}$ is $p(\cdot|\mu_0)$-Donsker, and thus $\Fc^2$ is $p(\cdot|\mu_0)$-Glivenko-Cantelli [see Lemma 2.10.4 in \cite{vandwell1996}]. Moreover, since $\Fc$ is $p(\cdot|\mu_0)$-Donsker so is $\Fc_{-} :=\{s_{\eta,\mu_0,-}|\eta \in \Gc^\eps\}$ [apply Theorem 2.10.6 in \cite{vandwell1996}], and thus $\Fc_{-}^2$ is also $p(\cdot|\mu_0)$-Glivenko-Cantelli. Hence we obtain
\[
\sup_{\eta \in \Gc} \Big|\frac{1}{n}\sum_{i=1}^n (s_{\eta,\mu_0}^2(X_{i}) - 1)\Big| + \Big|\frac{1}{n}\sum_{i=1}^n (s_{\eta,\mu_0,-}^2(X_{i}) - \|s_{\eta,\mu_0,-}\|^2_{2,\delta(\mu_0)})\Big|    = o_P(1). 
\]\
Thus
\[
\lim_{n \to \infty} \inf_{\eta \in \Gc\backslash\delta(\mu_0)} \frac{1}{n} \sum_{i=1}^n s_{ \eta,\mu_0,-}^2(X_{i}) \geq \inf_{\eta \in \Gc} \|s_{\eta,\mu_0,-}\|_{2,\delta(\mu_0)}^2> 0
\]
where the last inequality follows by the same arguments as (5) in \cite{gassiat2002}. Apply Inequality 1.1 from \cite{gassiat2002}, the lower bound above, and weak convergence of $\Gb_n$ to obtain
\begin{equation} \label{eq:proofhelp1_iid}
\sup_{\eta \in \Gc, \ell_n(\eta) - \ell_n(\delta(\mu_0))>0} \Big\|\frac{p_{\eta}}{p_{\delta(\mu_0)}} - 1\Big\|_{2,\delta(\mu_0)} = O_P(n^{-1/2}).
\end{equation}
Next, note that
\begin{equation} \label{eq:proof1_iid}
n^{-1} \sup_{\eta \in \Gc\backslash\delta(\mu_0)} \Big(\sum_{i=1}^n s_{\eta,
\mu_0}(X_{i}) \Big)^2 = \sup_{\eta \in \Gc} \Gb_n(\eta)^2 = O_P(1).
\end{equation}
The fact that $\Fc$ is Donsker and that $\E[s_{\eta,\mu_0}(X_i)] = 0$ implies that there must exist an envelope function $F$ of $\Fc$ with $\max_{i=1,..,n} F(X_i) = o_P(n^{1/2})$, this follows from Corollary 2.3.13 and Problem 2.3.4(iii) of \cite{vandwell1996}. Thus there exists $\alpha_n \to \infty$ such that $\sup_{i=1,...,n} F(X_{i}) = o_P(\alpha_n^{-1}n^{1/2})$. For such a sequence $\alpha_n$ define the sets
\[
M_{n1} := \{\eta \in \Gc: \ell_n(\eta) - \ell_n(\delta(\mu_0))>0\}, \quad M_{n2} := \Big\{\eta \in \Gc: 0 < \Big\|\frac{p_{\eta}}{p_{\delta(\mu_0)}} - 1\Big\|_{2,\delta(\mu_0)} \leq n^{-1/2}\alpha_n^{1/2} \Big\}.
\]
Note that
\begin{equation} \label{eq:proofhelp10_iid}
\sup_{\eta \in M_{n2}} \Big|\frac{1}{n}\sum_{i=1}^n (s_{\eta,\mu_0}^2(X_{i}) - 1)\Big| \leq \sup_{\eta \in \Gc} \Big|\frac{1}{n}\sum_{i=1}^n (s_{\eta,\mu_0}^2(X_{i}) - 1)\Big| = o_P(1).
\end{equation}
Now follow the arguments in the proof of Theorem \ref{th:main} which are used to obtain \eqref{eq:proof3} by replacing all instances of $\mu_n$ by $\mu_0$, all instances of $X_{i,n}$ by $X_i$,, all instances of $\ell_n^*$ by $\ell_n$ and using equations \eqref{eq:proofhelp1_iid}, \eqref{eq:proof1_iid} and \eqref{eq:proofhelp10_iid} instead of \eqref{eq:proofhelp1}, \eqref{eq:proof1} and \eqref{eq:proofhelp10} to arrive at the conclusion
\begin{equation} \label{eq:proof3_iid}
\sup_{\eta \in \bar \Gc} \ell_n(\eta) - \ell_n(\delta(\mu_0)) = \frac{1}{2}\sup_{\eta \in \Gc\backslash\delta(\mu_0)} \Big(\max\Big\{n^{-1/2}\sum_{i=1}^n s_{\eta,
\mu_0}(X_{i}), 0\Big\}\Big)^2 + o_P(1). 
\end{equation}
This proves \eqref{eq:iid1}, and the rest of the proof follows by a standard application of the multivariate CLT. \hfill $\Box$

\textbf{Proof of Theorem \ref{th:main}} The proof uses arguments from the proof of Theorem 3.1 in \cite{gassiat2002}. Let $\gamma_n := \|\mu_n - \mu_0\|$. Observe to each $\eta\in \Gc$ there exists $\tilde \eta \in \Gc^{\gamma_n}$ such that $\tilde\eta_n = \eta$. Thus under (A1) we have
\begin{equation} \label{eq:proof1}
n^{-1} \sup_{\eta \in \Gc\backslash\delta(\mu_n)} \Big(\sum_{i=1}^n s_{\eta,
\mu_n}(X_{i,n}) \Big)^2 \leq n^{-1} \sup_{\eta \in \Gc^{\gamma_n}} \Big(\sum_{i=1}^n s_{\eta_n,
\mu_n}(X_{i,n}) \Big)^2 \leq \sup_{\eta \in \Gc^{\eps}} \Gb_n^*(\eta)^2 = O_P(1)
\end{equation}
where the first inequality holds for $n$ sufficiently large. Moreover 
\[
\lim_{n \to \infty} \inf_{\eta \in \Gc\backslash\delta(\mu_n)} \frac{1}{n} \sum_{i=1}^n s_{ \eta,\mu_n,-}^2(X_{i,n}) \geq \lim_{n \to \infty} \inf_{\eta \in \Gc^{\gamma_n}} \frac{1}{n} \sum_{i=1}^n s_{ \eta_n,\mu_n,-}^2(X_{i,n}) \geq \inf_{\eta \in \Gc^{\eps}} \|s_{\eta,\mu_0,-}\|_{2,\delta(\mu_0)}^2> 0
\]
where the second inequality follows by (A2) and the third inequality follows by the same arguments as (5) in \cite{gassiat2002}. Apply Inequality 1.1 from \cite{gassiat2002} to obtain
\begin{equation} \label{eq:proofhelp1}
\sup_{\eta \in \Gc, \ell_n^*(\eta) - \ell_n^*(\delta(\mu_n))>0} \Big\|\frac{p_{\eta}}{p_{\delta(\mu_n)}} - 1\Big\|_{2,\delta(\mu_n)} = O_P(n^{-1/2}).
\end{equation}
By assumption (A3) there exist functions $F_n$ such that $\sup_{\eta \in \Gc} |s_{\eta,\mu_n}(x)| \leq F_n(x)$ and $\sup_{i=1,...,n} F_n(X_{i,n}) = o_P(n^{-1/2})$. Thus there exists $\alpha_n \to \infty$ such that $\sup_{i=1,...,n} F_n(X_{i,n}) = o_P(\alpha_n^{-1}n^{1/2})$. For such a sequence $\alpha_n$ define the sets
\[
M_{n1} := \{\eta \in \Gc: \ell_n^*(\eta) - \ell_n^*(\delta(\mu_n))>0\}, \quad M_{n2} := \Big\{\eta \in \Gc: 0 < \Big\|\frac{p_{\eta}}{p_{\delta(\mu_n)}} - 1\Big\|_{2,\delta(\mu_n)} \leq n^{-1/2}\alpha_n^{1/2} \Big\}.
\]
From \eqref{eq:proofhelp1} we obtain that $M_{n1} \subset M_{n2}$ with probability tending to one. On the other hand a Taylor expansion of $x \mapsto \log(1+x)$ shows that
\begin{align*}
& \sup_{\eta \in M_{n2}} \ell_n^*(\eta) - \ell_n^*(\delta(\mu_n))
\\
=& \  \sup_{\eta \in M_{n2}} \Big(\Big\|\frac{p_{\eta}}{p_{\delta(\mu_n)}} -1 \Big\|_{2,\delta(\mu_n)}\sum_{i=1}^n s_{\eta,
\mu_n}(X_{i,n}) - \frac{1}{2} \Big\|\frac{p_{\eta}}{p_{\delta(\mu_n)}} -1 \Big\|_{2,\delta(\mu_n)}^2 \sum_{i=1}^n s_{\eta,
\mu_n}^2(X_{i,n})
\\
& \ + \Big\|\frac{p_{\eta}}{p_{\delta(\mu_n)}} -1 \Big\|_{2,\delta(\mu_n)}^2 \sum_{i=1}^n s_{\eta,\mu_n}^2(X_{i,n}) R\Big(\Big\|\frac{p_{\eta}}{p_{\delta(\mu_n)}} -1 \Big\|_{2,\delta(\mu_n)} s_{\eta,\mu_n}(X_{i,n}) \Big)\Big)
\end{align*}
where the remainder function $R$ satisfies $R(u) \to 0$ for $u \to 0$. Now by the definition of $\alpha_n$ we have
\begin{align*}
& \sup_{\eta \in M_{n2}} \sum_{i=1}^n s_{\eta,\mu_n}^2(X_{i,n}) R\Big(\Big\|\frac{p_{\eta}}{p_{\delta(\mu_n)}} -1 \Big\|_{2,\delta(\mu_n)} s_{\eta,\mu_n}(X_{i,n}) \Big) 
\\
\leq &\ \sup_{\eta \in M_{n2}} \sum_{i=1}^n s_{\eta,\mu_n}^2(X_{i,n}) R\Big(n^{-1/2}\alpha_n^{1/2}o_P(\alpha_n^{-1}n^{1/2})\Big)
\\
= & \ o_P(1) \sup_{\eta \in M_{n2}} \sum_{i=1}^n s_{\eta,\mu_n}^2(X_{i,n}). 
\end{align*}
Additionally, (A2) implies that 
\begin{equation} \label{eq:proofhelp10}
\sup_{\eta \in M_{n2}} \Big|\frac{1}{n}\sum_{i=1}^n (s_{\eta,\mu_n}^2(X_{i,n}) - 1)\Big| \leq \sup_{\eta \in \Gc^\eps} \Big|\frac{1}{n}\sum_{i=1}^n (s_{\eta_n,\mu_n}^2(X_{i,n}) - 1)\Big| = o_P(1).
\end{equation}
Thus we see that 
\[
\sup_{\eta \in M_{n2}} \ell_n^*(\eta) - \ell_n^*(\delta(\mu_n)) = \sup_{\eta \in M_{n2}} \Big(\Big\|\frac{p_{\eta}}{p_{\delta(\mu_n)}} -1 \Big\|_{2,\delta(\mu_n)}\sum_{i=1}^n s_{\eta,
\mu_n}(X_{i,n}) - \frac{n}{2} \Big\|\frac{p_{\eta}}{p_{\delta(\mu_n)}} -1 \Big\|_{2,\delta(\mu_n)}^2 (1+r_n) \Big)
\]
where $r_n$ does not depend on $\eta$ and $r_n = o_P(1)$. Since $M_{n1} \subset M_{n2}$ with probability tending to one, and since 
\[
\sup_{\eta \in \bar \Gc} \ell_n^*(\eta) - \ell_n^*(\delta(\mu_n)) = \sup_{\eta \in M_{n1}} \ell_n^*(\eta) - \ell_n^*(\delta(\mu_n)), 
\]
it follows that
\begin{multline} 
\sup_{\eta \in \bar \Gc} \ell_n^*(\eta) - \ell_n^*(\delta(\mu_n)) = 
\\ \label{eq:proof2}
\sup_{\eta \in M_{n2}} \Big(\Big\|\frac{p_{\eta}}{p_{\delta(\mu_n)}} -1 \Big\|_{2,\delta(\mu_n)}\sum_{i=1}^n s_{\eta,
\mu_n}(X_{i,n}) - \frac{n}{2} \Big\|\frac{p_{\eta}}{p_{\delta(\mu_n)}} -1 \Big\|_{2,\delta(\mu_n)}^2 (1+r_n) \Big) + o_P(1).
\end{multline}
Next observe that under (A0), for any $\eta \in \Gc\backslash\delta(\mu_n)$ we also have $\eta^t := t\eta + (1-t)\delta(\mu_n) \in \Gc$ for any $t \in (0,1)$ provided that $\mu_n \in \Theta$. Additionally, we have
\[
\Big\|\frac{p_{\eta^t}}{p_{\delta(\mu_n)}} -1 \Big\|_{2,\delta(\mu_n)} = t \Big\|\frac{p_{\eta}}{p_{\delta(\mu_n)}} -1 \Big\|_{2,\delta(\mu_n)}
\]
and by construction $s_{\eta^t,\mu_n} \equiv s_{\eta,\mu_n}$.
Thus
\begin{align*}
& \sup_{\eta \in M_{n2}} \Big(\Big\|\frac{p_{\eta}}{p_{\delta(\mu_n)}} -1 \Big\|_{2,\delta(\mu_n)}\sum_{i=1}^n s_{\eta,
\mu_n}(X_{i,n}) - \frac{n}{2} \Big\|\frac{p_{\eta}}{p_{\delta(\mu_n)}} -1 \Big\|_{2,\delta(\mu_n)}^2 (1+r_n) \Big)
\\
= & \ \sup_{\eta \in \Gc \backslash\delta(\mu_n)} \sup_{0 < t\leq c_n(\eta)} \Big(t \Big\|\frac{p_{\eta}}{p_{\delta(\mu_n)}} - 1 \Big\|_{2,\delta(\mu_n)}\sum_{i=1}^n s_{\eta,
\mu_n}(X_{i,n}) - \frac{nt^2}{2} \Big\|\frac{p_{\eta}}{p_{\delta(\mu_n)}} -1 \Big\|_{2,\delta(\mu_n)}^2 (1+r_n) \Big) 
\end{align*} 
where $c_n(\eta) := n^{-1/2}\alpha_n^{1/2} \Big\|\frac{p_{\eta}}{p_{\delta(\mu_n)}} -1 \Big\|_{2,\delta(\mu_n)}^{-1}$. As soon as $r_n > -1$, which happens with probability tending to one, the supremum of the inner term over $t > 0$ is attained in the limit $t \to 0$ if $\sum_{i=1}^n s_{\eta, \mu_n}(X_{i,n}) \leq 0$ and at 
\[
t_n(\eta) := \frac{n^{-1} \sum_{i=1}^n s_{\eta, \mu_n}(X_{i,n})}{(1+r_n)\Big\|\frac{p_{\eta}}{p_{\delta(\mu_n)}} -1 \Big\|_{2,\delta(\mu_n)}}
\]
if $\sum_{i=1}^n s_{\eta, \mu_n}(X_{i,n}) > 0$. Because of \eqref{eq:proof1} it follows that $t_n(\eta) \leq c_n(\eta)$ with probability tending to one, so that taken together we have
\begin{align*}
& \sup_{\eta \in M_{n2}} \Big(\Big\|\frac{p_{\eta}}{p_{\delta(\mu_n)}} -1 \Big\|_{2,\delta(\mu_n)}\sum_{i=1}^n s_{\eta,
\mu_n}(X_{i,n}) - \frac{n}{2} \Big\|\frac{p_{\eta}}{p_{\delta(\mu_n)}} -1 \Big\|_{2,\delta(\mu_n)}^2 (1+r_n) \Big)
\\
= & \ \frac{1}{2(1+r_n)}\sup_{\eta \in \Gc\backslash\delta(\mu_n)} \Big(\max\Big\{n^{-1/2}\sum_{i=1}^n s_{\eta,
\mu_n}(X_{i,n}), 0\Big\}\Big)^2 + o_P(1) 
\\
= & \frac{1}{2}\sup_{\eta \in \Gc\backslash\delta(\mu_n)} \Big(\max\Big\{n^{-1/2}\sum_{i=1}^n s_{\eta,
\mu_n}(X_{i,n}), 0\Big\}\Big)^2 + o_P(1). 
\end{align*}
Combining this with \eqref{eq:proof2} yields
\begin{equation} \label{eq:proof3}
\sup_{\eta \in \bar \Gc} \ell_n^*(\eta) - \ell_n^*(\delta(\mu_n)) = \frac{1}{2}\sup_{\eta \in \Gc\backslash\delta(\mu_n)} \Big(\max\Big\{n^{-1/2}\sum_{i=1}^n s_{\eta,
\mu_n}(X_{i,n}), 0\Big\}\Big)^2 + o_P(1). 
\end{equation}
Recall that for each $\eta\in \Gc$ there exists $\tilde \eta \in \Gc^{\gamma_n}$ such that $\eta = \tilde \eta_n $. Thus
\begin{align} \nonumber
& \Big|\sup_{\eta \in \Gc\backslash\delta(\mu_n)} \Big(\max\Big\{n^{-1/2}\sum_{i=1}^n s_{\eta,
\mu_n}(X_{i,n}), 0\Big\}\Big)^2 - \sup_{\eta \in \Gc} \Big(\max\Big\{n^{-1/2}\sum_{i=1}^n s_{\eta_n,
\mu_n}(X_{i,n}), 0\Big\}\Big)^2\Big|
\\ \nonumber
\leq & \ \sup_{\nu \in \Gc\backslash\delta(\mu_n)} \inf_{\eta\in \Gc} \Big| \Big(n^{-1/2}\sum_{i=1}^n s_{\nu,\mu_n}(X_{i,n})\Big)^2 - \Big(n^{-1/2}\sum_{i=1}^n s_{\eta_n,\mu_n}(X_{i,n})\Big)^2 \Big|
\\ \nonumber
\leq & \ \sup_{\nu \in \Gc^{\gamma_n}} \inf_{\eta\in \Gc} \Big| \Big(n^{-1/2}\sum_{i=1}^n s_{\nu_n,\mu_n}(X_{i,n})\Big)^2 - \Big(n^{-1/2}\sum_{i=1}^n s_{\eta_n,\mu_n}(X_{i,n})\Big)^2 \Big|
\\ \nonumber
= & \ \sup_{\nu \in \Gc^{\gamma_n}} \inf_{\eta\in \Gc} \Big| (\Gb_n^*)^2(\nu) - (\Gb_n^*)^2(\eta) \Big|
\\  \label{eq:proof4}
\leq & \ 2\Big(\sup_{\nu \in \Gc^{\gamma_n}} |\Gb_n^*(\nu)|  \Big)\Big( \sup_{\nu \in \Gc^{\gamma_n}} \inf_{\eta\in \Gc} \Big| \Gb_n^*(\nu) - \Gb_n^*(\eta) \Big|\Big) = o_P(1)
\end{align}
The $o_P(1)$ in last line above follows from assumption (A1). More precisely, note that by the Continuous Mapping Theorem applied to the map $f \mapsto \sup_{\eta \in \Gc^\eps} \inf_{\tilde \eta \in \Gc} |f(\eta) - f(\tilde\eta)|$ we have for any fixed $\eps > 0$
\[
\sup_{\eta \in \Gc^\eps} \inf_{\tilde \eta \in \Gc} |\Gb_n^*(\eta) - \Gb_n^*(\tilde\eta)| \weak \sup_{\eta \in \Gc^\eps} \inf_{\tilde \eta \in \Gc} |\Gb^*(\eta) - \Gb^*(\tilde\eta)|. 
\]
Thus for arbitrary $\eps > 0, t > 0$ we have 
\[
\limsup_{n \to \infty} P\Big(\sup_{\eta \in \Gc^{\gamma_n}} \inf_{\tilde \eta \in \Gc} |\Gb_n^*(\eta) - \Gb_n^*(\tilde\eta)| \leq t \Big) \leq P\Big(\sup_{\eta \in \Gc^\eps} \inf_{\tilde \eta \in \Gc} |\Gb^*(\eta) - \Gb^*(\tilde\eta)| \leq t \Big),
\]
and the right-hand side can be made arbitrarily small by letting $\eps \downarrow 0$. This shows that
\[
\sup_{\nu \in \Gc^{\gamma_n}} \inf_{\eta\in \Gc} \Big| \Gb_n^*(\nu) - \Gb_n^*(\eta) \Big| = o_P(1).
\] 
Now equations \eqref{eq:proof3}, \eqref{eq:proof4} yield 
\[
2\sup_{\eta \in \bar \Gc} (\ell_n^*(\eta) - \ell_n^*(\delta(\mu_n))) = \sup_{\eta \in \Gc} \Big(\max\Big\{\Gb_n^*(\eta), 0\Big\}\Big)^2 + o_P(1), 
\]
and the first assertion of the theorem follows. The second assertion follows by an application of the continuous mapping theorem. \hfill $\Box$ \\

\textbf{Proof of Theorem \ref{th:continuity}} First we observe that $\Gb$ is the limit of $\Gb_n$ under weak convergence in $\ell^\infty(\Gc \setminus \delta(\mu_0) )$ and thus tight. Next, note that $\|Y\|^2 > 0$ almost surely. On the other hand, $L_n \geq 0$ almost surely for each $n$. Since $R$ is the weak limit of $2L_n$, it follows that $R \geq 0$ almost surely. Thus $\sup_{\eta} (\max\{\Gb(\eta), 0\})^2 > 0$ almost surely, and it follows  $\max(0,\sup_{\eta} \Gb(\eta)) = \sup_{\eta} \Gb(\eta) $ almost surely.

The proof of the first assertion [properties of $F_R$] consists of three steps. First, we show that the distribution of $R$ is continuous on $(0,\infty)$ (\textbf{Claim~2}). Second, we provide a lower bound for $P(R>0)$. Define
\[
F_y(t) := P\Big(\sup_{\eta} \Gb(\eta) \leq t \Big| Y_1 = y \Big).
\]
We begin by proving a preliminary result.\\
\\ 
\textbf{Claim 1}: \textit{For any $y \in \R^d$, $F_y(\cdot)$ is continuous on $(\|y\|,\infty)$.}\\
\\
Observe that by the joint normality of $(\Gb(\eta))_{\eta \in \Gc},Y$ the conditional distribution of $(\Gb(\eta))_{\eta \in \Gc}$ given $Y = y$ is that of a tight Gaussian random element with mean $\E[\Gb(\eta)Y^\top]y$ and a covariance function $\kappa$ that does not depend on $y$. Let $\tilde \Gb$ denote a centered Gaussian process with covariance function $\kappa$. Then the conditional distribution of $\Gb$ given $Y = y$ and the distribution of $(\tilde \Gb(\eta) + \E[\Gb(\eta)Y^\top]y)_{\eta \in \Gc}$ coincide.

Since $\tilde \Gb$ is a centered, tight Gaussian process, it follows by the arguments given on page 60-61 of \citeasnoun{LeTa1991} that $\sup_\eta |\tilde \Gb(\eta)|$ has a continuous distribution on $\R$ with left support point at $0$, so that $P(\sup_\eta |\tilde \Gb(\eta)|<\eps)>0$ for all $\eps>0$.  
Since $P(\sup_\eta \tilde \Gb(\eta) <\eps) \geq P(\sup_\eta |\tilde \Gb(\eta)|<\eps)$ it follows that also $P(\sup_\eta \tilde \Gb(\eta) <\eps) > 0$ for all $\eps  >0$. 

According to \citeasnoun{Ts1976}, the distribution of $\sup_\eta (\E[\Gb(\eta)Y^\top]y + \tilde \Gb(\eta))$ can only have a jump at the left endpoint of it's support and has a density to the right of that point. On the other hand, $|\E[\Gb(\eta)Y^\top]y| \leq \|\E[\Gb(\eta)Y]\|\|y\| \leq \|y\|$. Here, the second inequality follows since $\Gb(\eta),Y$ are jointly Gaussian so that there exist $a_\eta,b_\eta$ with $(\Gb(\eta),Y) \stackrel{\Dc}{=} (a_\eta^\top Y + b_\eta Z,Y)$ for $Z \sim \Nc(0,1)$ independent of $Y$. As $Y \sim \Nc(0,I_d)$ we have $1 = Var(\Gb(\eta)) = \|a_\eta\|^2 + b_\eta^2 \geq \|a_\eta\|^2$ and moreover $\|\E[\Gb(\eta)Y]\| = \|a_\eta\|$.

Thus for $\eps > 0,y \in \R^d$
\begin{align*}
P\Big(\sup_\eta \{\E[\Gb(\eta)Y^\top]y + \tilde \Gb(\eta)\} - \|y\| \leq \eps\Big) & = P\Big(\sup_\eta \{\E[\Gb(\eta)Y^\top]y - \|y\| + \tilde \Gb(\eta)\} \leq \eps\Big)
\\
& \geq P\Big(\sup_\eta \tilde \Gb(\eta) \leq \eps\Big) > 0. 
\end{align*} 
Thus for all $y \in \R^d$ the distribution of $\sup_\eta (\E[\Gb(\eta)Y^\top]y + \tilde \Gb(\eta))$ has a density on $(\|y\|,\infty)$ and Claim~1 follows. \\
\\
\textbf{Claim 2}: \textit{The distribution of $(\sup_\eta \Gb(\eta))^2 - \|Y\|^2$ is continuous on $(0,\infty)$}.\\
\\
Let $0<a<b$. Then by continuity of $F_y$ on $(\|y\|,\infty)$
\begin{align*}
P\Big( (\sup_{\eta} \Gb(\eta))^2 - \|Y\|^2 \in [a,b] \Big) &= \int_{\R} P\Big( (\sup_{\eta} \Gb(\eta))^2 - \|Y\|^2 \in [a,b] \Big| Y = y \Big)\phi_d(y) dy
\\
&= \int_{\R} \Big( F_y((\|y\|^2+b)^{1/2}) - F_y((\|y\|^2+a)^{1/2}) \Big) \phi_d(y) dy.
\end{align*}
Now for $a \uparrow b>0$ we have for every $y \in \R^d$ that $F_y((\|y\|^2+b)^{1/2}) - F_y((\|y\|^2+a)^{1/2}) \to 0$ since $(\|y\|^2+b)^{1/2} > \|y\|^2$ is a continuity point of $F_y$. Thus the integral converges to zero by dominated convergence. Since $b>0$ was arbitrary the assertion follows.\\
\\
\textbf{Claim 3}: \textit{For $d=1$ $P((\sup_\eta \Gb(\eta))^2 - Y^2> 0) \geq 1/4$}.\\
\\
By assumption there exists $\eta_0 \in \Gc$ such that $|\E[\Gb(\eta_0) Y]| \neq 1$. Moreover, 
\[
P((\sup_\eta \Gb(\eta))^2 - Y^2> 0) \geq P(|\Gb(\eta_0)| > |Y|) = 1/4. 
\]
Here, the last inequality follows since $(\Gb(\eta_0),Y)$ is a two-dimensional, centered Gaussian vector with $\E[(\Gb(\eta_0))^2] = \E[(Y)^2]$ and correlation in $(-1,1)$.\\

The continuity of $F_R$ on $(0,+\infty)$ and the bound $F_R(0) \leq 3/4$ in the case $d = 1$ follow by combining Claim~2 and Claim~3. 

It remains to establish the convergence $P(L_n > q_{n,1-\alpha}) \to \alpha$ in cases where $P(R>0) > \alpha$. Under the assumptions of the theorem, the maximum likelihood estimator $\hat \mu$ converges to $\mu_0$ in probability. Arguing along subsequences, we can without loss of generality assume that the convergence takes place almost surely. 
 
In what follows, denote by $\hat F_{n,B}$ the empirical distribution function of $L_{n,1},...,L_{n,B}$ and by $F_n$ the true distribution function of $L_{n,1}$ conditionally on $\hat \mu = \mu_n$. Note that conditionally on $\hat \mu = \mu_n$ the quantities $L_{n,1},...,L_{n,B}$ constitute an i.i.d. sample from $F_n$. By the uniform version of the Glivenko-Cantelli Theorem [see Theorem 2.8.1 in \cite{vandwell1996}] it follows that $\sup_{t\in\R} |\hat F_{n,B}(t) - F_n(t)| \to 0$ in probability, unconditionally. Additionally, the almost sure convergence $\hat \mu \to \mu_0$ together with Theorem~\ref{th:main} yields weak convergence of $L_{n,1}$ to $R$, so that $F_n$ converges to $F_R$ at all continuity points of $F_R$ almost surely. Thus we obtain that $\hat F_{n,B}$ converges to $F_R$ at all continuity points of $F_R$ in probability, and since $\hat F_{n,B},F_R$ are increasing and $F_R$ is continuous on $(0,\infty)$, $\sup_{x \in K} |\hat F_{n,B}(x) - F_R(x)|$ converges to zero in probability for compact $K \subset (0,\infty)$. By arguments similar to the ones given in Lemma 21.2 in \cite{vandervaart2000} we obtain that $\hat  q_{n,u} = \hat F_{n,B}^{-1}(u) \to F_R^{-1}(u)$ in probability for all $u$ where $F_R^{-1}$ is continuous. Note that $F_R^{-1}$ is increasing, and thus the set of its continuity points is dense in $[F_R(0),1]$. Moreover, $1-\alpha \in (F_R(0),1)$. Thus for every $\eps > 0$ there exist $1-\alpha_1 \leq 1-\alpha \leq 1-\alpha_2$ such that $F_R^{-1}$ is continuous at $1-\alpha_1,1-\alpha_2$ and $|\alpha_i - \alpha| \leq \eps$. By Slutzky's Lemma we obtain $L_n - \hat F_{n,B}^{-1}(1-\alpha_i) \weak R - F_R^{-1}(1-\alpha_i)$, and by continuity of $F_R$ in a  neighborhood of $F_R^{-1}(1-\alpha)$ and monotonicity of $\hat F_{n,B}^{-1}$ it follows that
\begin{align*}
1-\alpha_1 &= P(R - F_R^{-1}(1-\alpha_1)\leq 0) \leq \liminf_{n \to \infty} P(L_n \leq \hat  q_{n,1-\alpha}) \leq \limsup_{n \to \infty} P(L_n \leq \hat  q_{n,1-\alpha})
\\ 
&\leq  P(R - F_R^{-1}(1-\alpha_2)\leq 0) = 1-\alpha_2.
\end{align*}   
Since $\alpha_i$ above can be chosen to be arbitrarily close to $\alpha$ the claim follows. \hfill $\Box$ \\


\textbf{Proof of Proposition \ref{prop:loc-scale}} Note that the special structure of $p(\cdot|\mu)$ implies that $X_{1,n} \stackrel{\Dc}{=} X_1 - \mu_0 + \mu_n$ [recall that $X_{1,n} \sim p(\cdot|\mu_n), X_1 \sim p(\cdot|\mu_0)$]. On the other hand
\[
p_{\eta_n}(x) = \int p(x-\mu) d\eta_n(\mu) = \int p(x - \mu + \mu_0 - \mu_n) d\eta(\mu) = p_\eta(x + \mu_0 - \mu_n).
\]
Thus also $s_{\eta_n,\mu_n}(x) = s_{\eta,\mu}(x + \mu_0 - \mu_n)$ and in particular $s_{\eta_n,\delta(\mu_n)}(X_{1,n}) \stackrel{\Dc}{=} s_{\eta,\delta(\mu_0)}(X_{1})$. This in turn implies that for any measure $\eta \in\Gc^\eps$ we have by definition $\Gb_n^*(\eta)\stackrel{\Dc}{=} \Gb(\eta)$. Assuming that $\mu_0$ is an interior point of $\Theta$, similar computations show that $\ell'(X_{i,n}|\mu_n) \stackrel{\Dc}{=} \ell'(X_{i}|\mu_0)$ and $\|\ell'(\cdot|\mu_n)\|_{2,\delta(\mu_n)} = \|\ell'(\cdot|\mu_0)\|_{2,\delta(\mu_0)}$. Thus, the first part of (A1) follows. 

To verify assumption (A2), observe that $\Gb_n$ can be identified with the empirical process based on the observations $X_1,...,X_n$ and indexed by the class of functions $\Fc := \{s_{\eta,\mu_0}|\eta \in \Gc^\eps\}$. Weak convergence of $\Gb_n$ implies that the class $\Fc$ is $p(\cdot|\mu_0)$-Donsker, and thus $\Fc^2$ is $p(\cdot|\mu_0)$-Glivenko-Cantelli [see Lemma 2.10.4 in \cite{vandwell1996}]. Moreover, since $\Fc$ is $p(\cdot|\mu_0)$-Donsker so is $\Fc_{-} :=\{s_{\eta,\mu_0,-}|\eta \in \Gc^\eps\}$ [apply Theorem 2.10.6 in \cite{vandwell1996}], and thus $\Fc_{-}^2$ is also $p(\cdot|\mu_0)$-Glivenko-Cantelli. This shows that (A2) holds.

For assumption (A3), note that for every $\eta \in \Gc$ there exists $\tilde \eta \in \Gc^\eps$ with $\tilde \eta_n = \eta$ provided that $\|\mu_n-\mu_0\| \leq \eps$. Thus
$s_{\eta_n,\mu_n}(x) = s_{\eta,\mu_0}(x+\mu_0-\mu_n)$ implies that for any $x \in \R$
\[
\sup_{f \in \Fc_n} |f(x)| \leq \sup_{\eta \in \Gc^\eps} |s_{\eta_n,\mu_n}(x)| = \sup_{\eta \in \Gc^\eps} |s_{\eta,\mu_0}(x + \mu_0 - \mu_n)|. 
\]
Thus if $F$ is an envelope for $\Fc^\eps := \{s_{\eta,\mu_0}|\eta \in \Gc^\eps\}$ then $F_n(\cdot) := F(\cdot + \mu_0 - \mu_n)$ is an envelope for $\Fc_n$. On the other hand, the fact that $\Fc^\eps$ is Donsker and that $\E[s_{\eta,\mu_0}(X_i)] = 0$ implies that there must exist an envelope function $F$ of $\Fc^\eps$ with $\max_{i=1,..,n} F(X_i) = o_P(n^{1/2})$, this follows from Corollary 2.3.13 and Problem 2.3.4(iii) of \cite{vandwell1996}. Moreover, $F_n(X_{i,n}) \stackrel{\Dc}{=} F(X_i)$ and thus (A3) follows. \hfill $\Box$

\section{Verification of Assumptions (A1) - (A3) for Poisson Mixtures} \label{sec:poisdet}

Assume that $\Theta = [a,b]$ for some $0<a<b$ and that the densities $p$ take the form $p(x|\mu) = \mu^x e^{-\mu}/x!$ with respect to the counting measure on $\N$. As stated in Section 3.3 of \cite{Azgame2009}, the likelihood ratios have the following representation
\begin{equation} \label{eq:ak}
\frac{p_{\eta_n}(x)}{p_{\delta(\mu_n)}(x)} - 1 = \sum_{k=1}^\infty \frac{k \E[(Z-\mu_n)^k]}{(k!\mu_n^k)^{1/2}} \frac{C_k(x|\mu_n)}{k} =: \sum_{k=1}^\infty a_k(\eta_n,\mu_n) \frac{C_k(x|\mu_n)}{k}
\end{equation}
where $Z \sim \eta_n$. Here, the functions $x \mapsto C_k(x|\mu_n)$ are polynomials of order $k$ which are given by
\[
C_k(x|\mu_n) := \frac{\mu_n^{k/2}}{(k!)^{1/2}} \Big[\frac{d^k}{dz^k} \Big(\frac{z}{\mu_n}\Big)^x\exp(-z+\mu_n)\Big]_{z = \mu_n}.
\]
The functions $(x \mapsto C_k(x|\mu_n))_{k \in \N}$ are centered and orthonormal with respect to $P_{\delta(\mu_n)}$, i.e. for $k,\ell\in\N$
\begin{equation} \label{eq:ak2}
\E[C_k(X_{1,n}|\mu_n)] = 0, \quad \E[C_k(X_{1,n}|\mu_n)C_\ell(X_{1,n}|\mu_n)] =  I\{k=\ell\}.
\end{equation}
In particular, we have that 
\[
1 = \E[C_k^2(X_{1,n}|\mu_n)] = \sum_{u \geq 0} C_k^2(u|\mu_n) e^{-\mu_n}\mu_n^u/u!\geq C_k^2(x|\mu_n) e^{-\mu_n}\mu_n^x/x! \quad \forall\ x\in\N_0
\]
so that the series in \eqref{eq:ak} converges pointwise.
The score functions $s_{\eta_n,\mu_n}$ can be represented as
\begin{equation} \label{eq:scorespois}
s_{\eta_n,\mu_n}(x) = \sum_{k=1}^\infty  \frac{a_k(\eta_n,\mu_n)C_k(x|\mu_n)}{k w(\eta_n,\mu_n)}, \quad w(\eta_n,\mu_n) := \Big(\sum_{\ell=1}^\infty \ell^{-2}a_\ell^2(\eta_n,\mu_n) \Big)^{1/2}.
\end{equation} 
For $L \geq 2$, define the approximating function
\[
s_{\eta_n,\mu_n}^{(L)}(x) = \sum_{k=1}^L  \frac{a_k(\eta_n,\mu_n)C_k(x|\mu_n)}{k w^{(L)}(\eta_n,\mu_n)}, \quad w^{(L)}(\eta_n,\mu_n) := \Big(\sum_{\ell=1}^L \ell^{-2}a_\ell^2(\eta_n,\mu_n) \Big)^{1/2}.
\] 
Obviously, the function $x \mapsto s_{\eta_n,\mu_n}^{(L)}(x)$ is a polynomial of degree $L$. Later, we will prove the following identities holding for $L \geq 2$, some finite $n_0$ and a constant $C$ independent of $n,\eta_n,\mu_n,\mu_0$
\begin{align} \label{eq:phelp1}
\sup_{\eta \in \Gc^\eps}\sup_{n \geq n_0}\Big|\frac{w^{(L)}(\eta_n,\mu_n)}{w(\eta_n,\mu_n)} - 1\Big| &\leq CL^{-1}, \quad \sup_{\eta \in \Gc^\eps} \Big|\frac{w^{(L)}(\eta,\mu_0)}{w(\eta,\mu_0)} - 1\Big| \leq CL^{-1},
\\ \label{eq:phelp2}
\sum_{k \geq 2} a_k^2(\eta_n,\mu_n) &\leq C a_2^2(\eta_n,\mu_n).
\end{align}
Additionally, for any fixed $k$ one obtains by straightforward calculations
\begin{equation} \label{eq:phelp3}
\sup_{\eta \in \Gc^\eps} |a_k(\eta_n,\mu_n) - a_k(\eta,\mu_0)| \to 0, \quad n \to \infty,
\end{equation}
and for any fixed $L \geq 2$ [this will be proved later]
\begin{equation} \label{eq:phelp4}
\sup_{\eta \in \Gc^\eps}\Big|\frac{w^{(L)}(\eta_n,\mu_n)}{w^{(L)}(\eta,\mu_0)} - 1\Big| \to 0, \quad n \to \infty.
\end{equation}

Assumption (A3) can be verified by a straightforward extension of the arguments in the proof of Theorem 4 of \cite{Azgame2009}. Details are omitted for the sake of brevity. In the proofs that follow, we will repeatedly use (A3).\\
\\
\textbf{Verification of Assumption (A1)}. To establish assertion (A1), it suffices to prove asymptotic tightness of the process $\Gb_n^*$ in $\ell^\infty(\Gc_\eps)$ and that weak convergence
\[
\Big(\Gb_n^*(\eta_1),...,\Gb_n^*(\eta_k),\frac{1}{\sqrt{n}} \sum_{i=1}^{n} \|\ell'(\cdot|\mu_0)\|_{2,\delta(\mu_0)}^{-1}\ell'(X_{i,n}|\mu_n)\Big) \weak (\Gb(\eta_1),...,\Gb(\eta_k),Y_1)
\] 
holds for any fixed collection of measures $\eta_1,....\eta_k$. The weak convergence above follows by straightforward arguments, and we will only provide the details for establishing tightness. To prove asymptotic tightness of $\Gb_n^*$, we will prove that $\Gb_n^* \weak \Gb$. For $L\geq 2$ define
\[
\Gb^{(L)}(\eta) := \sum_{k=1}^L \frac{a_k(\eta,\mu_0)Z_k}{k w^{(L)}(\eta,\mu_0)}, \quad \Gb(\eta) := \sum_{k=1}^\infty  \frac{a_k(\eta,\mu_0)Z_k}{k w(\eta,\mu_0)}
\] 
where $Z_1,Z_2,...$ i.i.d. $\sim \Nc(0,1)$. In what follows, define for an arbitrary function $f: \R \to \R$ with $\E|f(X_{1,n})|<\infty$
\[
\Fb_n f := \frac{1}{n^{1/2}} \sum_{i=1}^n (f(X_{i,n}) - \E[f(X_{i,n})]).
\]
Note that by construction $\Gb_n^*(\eta) = \Fb_n s_{\eta_n,\mu_n}$. By an application of Lemma B.1 from \cite{budevo2011}, weak convergence of $\Gb_n^*$ to $\Gb$ follows from the following three claims:
\begin{enumerate}
\item[(i)] For every $L \geq 2$ we have $(\Fb_n s_{\eta_n,\mu_n}^{(L)})_{\eta \in \Gc^\eps} \weak (\Gb^{(L)})_{\eta \in \Gc^\eps}$ as $n \to \infty$.
\item[(ii)] $\Gb^{(L)} \weak \Gb$ as $L \to \infty$.
\item[(iii)] For every $\delta >0$ we have [with $P^*$ denoting outer probability]
\[
\lim_{L \to \infty} \limsup_{n \to \infty} P^*\Big(\sup_{\eta \in \Gc^\eps} |\Fb_n s_{\eta_n,\mu_n}^{(L)}-\Fb_n s_{\eta_n,\mu_n}| > \delta \Big) = 0.
\]
\end{enumerate} 
For a proof of (iii) note that
\begin{align*}
\Fb_n s_{\eta_n,\mu_n}-\Fb_n s_{\eta_n,\mu_n}^{(L)}  = & \Big(1 - \frac{w(\eta_n,\mu_n)}{w^{(L)}(\eta_n,\mu_n)} \Big)\sum_{k=1}^\infty \frac{a_k(\eta_n,\mu_n)}{kw(\eta_n,\mu_n)} \Fb_n C_k(\cdot|\mu_n)
\\
& + \frac{w(\eta_n,\mu_n)}{w^{(L)}(\eta_n,\mu_n)} \sum_{k=L+1}^\infty \frac{a_k(\eta_n,\mu_n)}{kw(\eta_n,\mu_n)} \Fb_n C_k(\cdot|\mu_n)
\\
=: & A_{n}^{(L)}(\eta_n,\mu_n) + B_{n}^{(L)}(\eta_n,\mu_n).
\end{align*}
The first term in the above decomposition can be bounded as follows
\begin{align*}
\sup_{\eta \in \Gc^\eps} |A_{n}^{(L)}(\eta_n,\mu_n)| = &  \sup_{\eta \in \Gc^\eps} \Big|\Big(1 - \frac{w(\eta_n,\mu_n)}{w^{(L)}(\eta_n,\mu_n)} \Big)\sum_{k=1}^\infty \frac{a_k(\eta_n,\mu_n)}{kw(\eta_n,\mu_n)} \Fb_n C_k(\cdot|\mu_n) \Big|
\\
\leq  &
CL^{-1}\Big(\sum_{k=1}^\infty  \frac{(\Fb_n C_k(\cdot|\mu_n))^2}{k^2} \Big)^{1/2}\sup_{\eta \in \Gc^\eps }\Big(\sum_{k=1}^\infty \frac{a_k^2(\eta_n,\mu_n)}{w^2(\eta_n,\mu_n)} \Big)^{1/2} 
\\
\leq &
CL^{-1} \Big(\sum_{k=1}^\infty \frac{(\Fb_n C_k(\cdot|\mu_n))^2}{k^2} \Big)^{1/2}\sup_{\eta \in \Gc^\eps }\Big( \frac{a_1^2(\eta_n,\mu_n) + C a_2^2(\eta_n,\mu_n)}{a_1^2(\eta_n,\mu_n) + a_2^2(\eta_n,\mu_n)/4} \Big)^{1/2}
\\
\leq & \tilde CL^{-1} \Big(\sum_{k=1}^\infty \frac{(\Fb_n C_k(\cdot|\mu_n))^2}{k^2} \Big)^{1/2},
\end{align*}
where the first inequality follows from \eqref{eq:phelp1} and the second inequality from \eqref{eq:phelp2}. Since $\E[(\Fb_n C_k(\cdot|\mu_n))^2] = 1$ for all $k \in \N$ by the orthonormality of the $(C_k(\cdot|\mu_n))_{k \in \N}$, we obtain
\[
\lim_{L \to \infty} \limsup_{n \to \infty} \E\Big| \sup_{\eta \in \Gc^\eps} A_{n}^{(L)}(\eta_n,\mu_n) \Big|^2 = 0. 
\]
By similar arguments as above we also obtain the bound
\begin{align*}
\sup_{\eta \in \Gc^\eps} |B_{n}^{(L)}(\eta_n,\mu_n)| & \leq C_1 \Big(\sum_{k=L+1}^\infty \frac{(\Fb_n C_k(\cdot|\mu_n))^2}{k^2} \Big)^{1/2} \sup_{\eta \in \Gc^\eps}\Big( \frac{w(\eta_n,\mu_n)}{w^{(L)}(\eta_n,\mu_n)}\Big) 
\\
&\leq C_2 \Big(\sum_{k=L+1}^\infty \frac{(\Fb_n C_k(\cdot|\mu_n))^2}{k^2} \Big)^{1/2}
\end{align*}
where the last inequality holds for $n$ sufficiently large. Thus
\[
\lim_{L \to \infty} \limsup_{n \to \infty} \E\Big| \sup_{\eta \in \Gc^\eps} B_{n}^{(L)}(\eta_n,\mu_n) \Big|^2 \leq \lim_{L \to \infty}C_2 \sum_{k = L+1}^\infty \frac{1}{k^2} = 0.
\]
and assertion (iii) follows. Assertion (ii) can be proved by similar arguments with $Z_k$ replacing $\Fb_n C_k(\cdot|\mu_n)$ and the arguments are omitted for brevity. For the proof of assertion (i), note that for any fixed $L$ it is easy to verify that 
\[
(\Fb_n C_1(\cdot|\mu_n),...,\Fb_n C_L(\cdot|\mu_n)) \weak (Z_1,...,Z_L).
\]
To see this, recall that the $C_k(\cdot|\mu_n)$ are polynomials and that for $\mu_n \to \mu_0$ the coefficients of $C_k(\cdot|\mu_n)$ converge to those of $C_k(\cdot|\mu_0)$. Weak convergence of $(\Fb_n s_{\eta_n,\mu_n}^{(L)})_{\eta \in \Gc^\eps}$ follows by the extended continuous mapping theorem [see Theorem 1.11.1 in \cite{vandwell1996}] applied to the maps [to verify the conditions of the continuous mapping theorem, make use \eqref{eq:phelp3}-\eqref{eq:phelp4}]
\[
g_n: (x_1,...,x_L) \mapsto \Big(\sum_{k=1}^L \frac{a_k(\eta_n,\mu_n)x_k}{k w^{(L)}(\eta_n,\mu_n)} \Big)_{\eta \in \Gc^\eps}, \quad g: (x_1,...,x_L) \mapsto \Big(\sum_{k=1}^L \frac{a_k(\eta,\mu_0)x_k}{k w^{(L)}(\eta,\mu_0)} \Big)_{\eta \in \Gc^\eps}.
\] 
Thus (i)-(iii) are established and we see that weak convergence of $\Gb_n$ holds and the limiting Gaussian process $\Gb$ has the following covariance structure (this follows after some calculations)
\[
\E[\Gb(\eta_1)\Gb(\eta_2)] = \frac{\E[\exp((Z_1-\mu)(Z_2-\mu)/\mu)] - 1}{(\E[\exp((Z_1-\mu)(\tilde Z_1-\mu)/\mu)] - 1)^{1/2}(\E [\exp((Z_2-\mu)(\tilde Z_2-\mu)/\mu)] - 1)^{1/2}} 
\]
where $Z_1,\tilde Z_1 \sim\eta_1, Z_2,\tilde Z_2 \sim \eta_2$ and $Z_1,Z_2,\tilde Z_1, \tilde Z_2$ are independent. Equation \eqref{eq:epsto0*} can be proved by arguments similar to those in Example~\ref{ex:gauss}. Thus we have established (A1).\\
\\
\textbf{Verification of condition (A2)}. Consider the following decomposition
\begin{align}
& \E \sup_{\eta \in \Gc^\eps} \Big|\frac{1}{n}\sum_{i=1}^n s_{\eta_n,\mu_n}^2(X_{i,n}) - (s_{\eta_n,\mu_n}^{(L)})^2(X_{i,n})\Big| \nonumber
\\
= & \E \sup_{\eta \in \Gc^\eps}\Big|\frac{1}{n}\sum_{i=1}^n [s_{\eta_n,\mu_n}(X_{i,n}) - s_{\eta_n,\mu_n}^{(L)}(X_{i,n})][s_{\eta_n,\mu_n}(X_{i,n}) + s_{\eta_n,\mu_n}^{(L)}(X_{i,n})]\Big| \nonumber
\\
\leq & \E \Big[\Big(\sup_{\eta \in \Gc^\eps}\frac{1}{n}\sum_{i=1}^n [s_{\eta_n,\mu_n}(X_{i,n}) - s_{\eta_n,\mu_n}^{(L)}(X_{i,n})]^2 \Big)^{1/2} \nonumber
\\
& \quad \times \Big(\sup_{\eta \in \Gc^\eps}\frac{1}{n}\sum_{i=1}^n [s_{\eta_n,\mu_n}(X_{i,n}) + s_{\eta_n,\mu_n}^{(L)}(X_{i,n})]^2 \Big)^{1/2}\Big] \nonumber
\\
\leq &  \E \Big[\sup_{\eta \in \Gc^\eps}\frac{1}{n}\sum_{i=1}^n [s_{\eta_n,\mu_n}(X_{i,n}) - s_{\eta_n,\mu_n}^{(L)}(X_{i,n})]^2 \Big] \E \Big[\sup_{\eta \in \Gc^\eps}\frac{1}{n}\sum_{i=1}^n [s_{\eta_n,\mu_n}(X_{i,n}) + s_{\eta_n,\mu_n}^{(L)}(X_{i,n})]^2 \Big]. \label{eq:pogc1}
\end{align}
Moreover, for $n$ sufficiently large and some constants $C_2,\tilde C$ we obtain by arguments similar to the ones in the proof of 
\begin{align*}
& \sup_{\eta \in \Gc^\eps}|s_{\eta_n,\mu_n}(X_{i,n}) - s_{\eta_n,\mu_n}^{(L)}(X_{i,n})|
\\
\leq & \sup_{\eta \in \Gc^\eps}\Big|1 - \frac{w(\eta_n,\mu_n)}{w^{(L)}(\eta_n,\mu_n)} \Big| \Big|\sum_{k=1}^\infty \frac{a_k(\eta_n,\mu_n)}{kw(\eta_n,\mu_n)} C_k(X_{i,n}|\mu_n)\Big|
\\
& + \sup_{\eta \in \Gc^\eps}\Big|\frac{w(\eta_n,\mu_n)}{w^{(L)}(\eta_n,\mu_n)}\Big| \Big|\sum_{k=L+1}^\infty \frac{a_k(\eta_n,\mu_n)}{kw(\eta_n,\mu_n)} C_k(X_{i,n}|\mu_n)\Big|
\\
\leq & \sup_{\eta \in \Gc^\eps}\Big|1 - \frac{w(\eta_n,\mu_n)}{w^{(L)}(\eta_n,\mu_n)} \Big| \Big|\sum_{k=1}^\infty \frac{a_k^2(\eta_n,\mu_n)}{w^2(\eta_n,\mu_n)}\Big|^{1/2}  \Big|\sum_{k=1}^\infty \frac{C_k^2(X_{i,n}|\mu_n)}{k^2}\Big|^{1/2}
\\
& + \sup_{\eta \in \Gc^\eps}\Big|\frac{w(\eta_n,\mu_n)}{w^{(L)}(\eta_n,\mu_n)}\Big|   \Big|\sum_{k=L+1}^\infty \frac{a_k^2(\eta_n,\mu_n)}{w^2(\eta_n,\mu_n)}\Big|^{1/2}  \Big|\sum_{k=L+1}^\infty \frac{C_k^2(X_{i,n}|\mu_n)}{k^2}\Big|^{1/2}
\\
\leq & \tilde CL^{-1} \Big|\sum_{k=1}^\infty \frac{C_k^2(X_{i,n}|\mu_n)}{k^2}\Big|^{1/2} + C_2 \Big|\sum_{k=L+1}^\infty \frac{C_k^2(X_{i,n}|\mu_n)}{k^2}\Big|^{1/2} 
\end{align*}
where the last inequality follows from \eqref{eq:phelp1} and \eqref{eq:phelp2}. The last identity shows that for some constant $C_3$ and $n$ sufficiently large
\begin{equation} \label{eq:pogc2}
\E\sup_{\eta \in \Gc^\eps}|s_{\eta_n,\mu_n}(X_{i,n}) - s_{\eta_n,\mu_n}^{(L)}(X_{i,n})|^2 \leq C_3 \Big(L^{-2} + \sum_{k=L+1}^\infty \frac{1}{k^2} \Big).  
\end{equation}
Combining (A3) with \eqref{eq:pogc1} and \eqref{eq:pogc2} shows that
\begin{equation} \label{eq:pogc3}
\limsup_{n \to \infty}\E \sup_{\eta \in \Gc^\eps} \Big|\frac{1}{n}\sum_{i=1}^n s_{\eta_n,\mu_n}^2(X_{i,n}) - (s_{\eta_n,\mu_n}^{(L)})^2(X_{i,n})\Big| \leq C_4 \Big(L^{-2} + \sum_{k=L+1}^\infty \frac{1}{k^2} \Big).  
\end{equation}
Next, observe that by construction we have $\E[(s_{\eta_n,\mu_n}^{(L)})^2(X_{i,n})] = 1$ for all $n \in \N, L\geq 2, \eta \in \Gc^\eps$. Moreover simple arguments show that for every fixed $k,l \in \N$
\[
\frac{1}{n}\sum_{i=1}^n C_k(X_{i,n}|\mu_n)C_l(X_{i,n}|\mu_n) \stackrel{P}{\to} I\{k=l\}.
\]
By the extended continuous mapping theorem [see Theorem 1.11.1 in \cite{vandwell1996}] applied to the maps 
\begin{align*}
&g_n: (x_{kl})_{k,l=1,...,L} \mapsto \Big(\sum_{k,l=1}^L  \frac{a_k(\eta_n,\mu_n)a_l(\eta_n,\mu_n) x_{kl}}{kl (w^{(L)}(\eta_n,\mu_n))^2} \Big)_{\eta \in \Gc^\eps}
\\
&g: (x_{kl})_{k,l=1,...,L} \mapsto \Big(\sum_{k,l=1}^L  \frac{a_k(\eta,\mu_0)a_l(\eta,\mu_0)x_{kl}}{kl (w^{(L)}(\eta,\mu_0))^2} \Big)_{\eta \in \Gc^\eps}
\end{align*}
it follows that for every $L\geq 2$
\[
\sup_{\eta \in \Gc^\eps} \Big|\frac{1}{n}\sum_{i=1}^n ( (s_{\eta_n,\mu_n}^{(L)})^2(X_i) - 1)\Big| = o_P(1).
\]
Combining this with \eqref{eq:pogc3} proves the first part of assertion (A2). To establish the second part of (A2), note that for $x,y \in \R$ we have $|x_- -y_-| \leq |x-y|$. Thus
\begin{align*}
& \sup_{\eta \in \Gc^\eps} \Big|\frac{1}{n}\sum_{i=1}^n s_{\eta_n,\mu_n,-}^2(X_{i,n}) - (s_{\eta_n,\mu_n,-}^{(L)})^2(X_{i,n})\Big|
\\
\leq & \sup_{\eta \in \Gc^\eps} \Big|\frac{1}{n}\sum_{i=1}^n (s_{\eta_n,\mu_n,-}(X_{i,n}) - s_{\eta_n,\mu_n,-}^{(L)}(X_{i,n}))^2\Big|^{1/2}\Big|\frac{1}{n}\sum_{i=1}^n (s_{\eta_n,\mu_n,-}(X_{i,n}) + s_{\eta_n,\mu_n,-}^{(L)}(X_{i,n}))^2\Big|^{1/2}
\\
\leq & \sup_{\eta \in \Gc^\eps} \Big\{ \Big|\frac{1}{n}\sum_{i=1}^n (s_{\eta_n,\mu_n}(X_{i,n}) - s_{\eta_n,\mu_n}^{(L)}(X_{i,n}))^2\Big|^{1/2}
\\
& \quad\quad \times \Big|\frac{4}{n}\sum_{i=1}^n 4(s_{\eta_n,\mu_n}(X_{i,n}))^2 + (s_{\eta_n,\mu_n}(X_{i,n})-s_{\eta_n,\mu_n}^{(L)}(X_{i,n}))^2\Big|^{1/2} \Big\}.
\end{align*}
This combined with \eqref{eq:pogc2} and (A3) yields
\begin{equation} \label{eq:pa20}
\limsup_{n \to \infty}\E \sup_{\eta \in \Gc^\eps} \Big|\frac{1}{n}\sum_{i=1}^n s_{\eta_n,\mu_n,-}^2(X_{i,n}) - (s_{\eta_n,\mu_n,-}^{(L)})^2(X_{i,n})\Big| \leq C_4 \Big(L^{-2} + \sum_{k=L+1}^\infty \frac{1}{k^2} \Big).  
\end{equation}
Thus it suffices to show that for each fixed $L$
\begin{equation} \label{eq:pa21}
\sup_{\eta \in \Gc^\eps} \Big|\frac{1}{n}\sum_{i=1}^n (s_{\eta_n,\mu_n,-}^{(L)})^2(X_{i,n}) - \|s_{\eta_n,\mu_n,-}^{(L)}\|_{2,\delta(\mu_n)}^2\Big| = o_P(1)
\end{equation}
and that
\begin{equation} \label{eq:pa22}
\lim_{L \to \infty} \limsup_{n \to \infty}\sup_{\eta \in \Gc^\eps} \Big|\|s_{\eta_n,\mu_n,-}^{(L)}\|_{2,\delta(\mu_n)}^2 - \|s_{\eta,\mu_0,-}\|_{2,\delta(\mu_0)}^2\Big| = 0.
\end{equation}
To prove \eqref{eq:pa21}, define $y^{(L)}(x):=(1,...,x^L)$ and observe that there exists a constant $C$ [note that $s_{\eta_n,\mu_n}^{(L)}(x)$ is a polynomial in $x$ of degree $L$] such that
\begin{multline*}
\sup_{\eta \in \Gc^\eps} \Big|\frac{1}{n}\sum_{i=1}^n (s_{\eta_n,\mu_n,-}^{(L)})^2(X_{i,n}) - \|s_{\eta_n,\mu_n,-}^{(L)}\|_{2,\delta(\mu_n)}^2\Big|
\\ 
\leq \sup_{b \in \R^{L+1}, \|b\| \leq C} \Big| \frac{1}{n}\sum_{i=1}^n (b^T Y^{(L)}(X_{i,n}))^2I\{b^T Y^{(L)}(X_{i,n}) \leq 0\}
\\
- \E[(b^T Y^{(L)}(X_{i,n}))^2I\{b^T Y^{(L)}(X_{i,n}) \leq 0\}] \Big|.
\end{multline*}
Weak convergence to zero of the right-hand side can be proved after observing that the class of functions $\{y \mapsto (b^T y)^2I\{b^T y \leq 0\} : \|b\| \leq C\}$ is VC and has an envelope $G$ function which satisfies $\sup_{n \geq n_0} \E G^2(Y^{(L)}(X_{i,n})) < \infty$ for some $n_0 < \infty$. Thus convergence of the right-hand side above to zero follows from Theorem 2.8.1 in \cite{vandwell1996}. 

Next, let us prove \eqref{eq:pa22}. We begin by proving
\begin{equation} \label{eq:pa23}
\limsup_{n \to \infty}\sup_{\eta \in \Gc^\eps} \Big|\|s_{\eta_n,\mu_n,-}^{(L)}\|_{2,\delta(\mu_n)}^2 - \|s_{\eta,\mu_0,-}^{(L)}\|_{2,\delta(\mu_n)}^2\Big| + \Big|\|s_{\eta,\mu_0,-}^{(L)}\|_{2,\delta(\mu_n)}^2 - \|s_{\eta,\mu_0,-}^{(L)}\|_{2,\delta(\mu_0)}^2\Big| = 0
\end{equation}
for every fixed $L \geq 2$. Convergence to zero of $\sup_{\eta \in \Gc^\eps} \Big|\|s_{\eta_n,\mu_n,-}^{(L)}\|_{2,\delta(\mu_n)}^2 - \|s_{\eta,\mu_0,-}^{(L)}\|_{2,\delta(\mu_n)}^2\Big|$ follows from the fact that, for $V_n \sim Pois(\mu_n)$, we have for some sequence $\alpha_n = o(1)$
\begin{align} \nonumber
& \ \sup_{\eta \in \Gc^\eps} \Big|\|s_{\eta_n,\mu_n,-}^{(L)}\|_{2,\delta(\mu_n)}^2 - \|s_{\eta,\mu_0,-}^{(L)}\|_{2,\delta(\mu_n)}^2\Big|
\\ \nonumber
\leq & \ \sup_{\|a-b\|\leq \alpha_n, \|a\| \leq C, \|b\|\leq C} \E\Big|(b^T Y^{(L)}(V_n))^2I\{b^T Y^{(L)}(V_n) \leq 0\}
\\
& \ \quad\quad\quad\quad -(a^T Y^{(L)}(V_n))^2I\{a^T Y^{(L)}(V_n) \leq 0\} \Big| \nonumber
\\ \label{eq:pa24}
\leq & \ 2 C \alpha_n \E[\|Y^{(L)}(V_n)\|^4] = o(1)
\end{align}
where the last inequality follows from $|x_-^2-y_-^2| \leq (|x|+|y|)(|x|-|y|)$. Similarly, letting $V_0 \sim Pois(\mu_0)$, the second term can be bounded by
\begin{align*}
& \ \sup_{\eta \in \Gc^\eps} \Big|\|s_{\eta,\mu_0,-}^{(L)}\|_{2,\delta(\mu_n)}^2 - \|s_{\eta,\mu_0,-}^{(L)}\|_{2,\delta(\mu_0)}^2\Big|
\\
\leq &\ \sup_{b \in \R^{L+1}, \|b\| \leq C}  \Big| \E[(b^T Y^{(L)}(V_n))^2I\{b^T Y^{(L)}(V_n) \leq 0\}] - \E[(b^T Y^{(L)}(V_0))^2I\{b^T Y^{(L)}(V_0) \leq 0\}] \Big|.
\end{align*}
Covering $B := \{b \in \R^{L+1}: \|b\| \leq C\}$ with a finite number of balls of radius $\eps$ one can reduce the above problem to showing that
\[
\E[(b^T Y^{(L)}(V_n))^2I\{b^T Y^{(L)}(V_n) \leq 0\}] \to \E[(b^T Y^{(L)}(V_0))^2I\{b^T Y^{(L)}(V_0) \leq 0\}] 
\]
for any fixed $b \in B$. Observe that $V_n$ converges weakly to $V$. The continuous mapping theorem implies that $(b^T Y^{(L)}(V_n))^2I\{b^T Y^{(L)}(V_n) \leq 0\} \weak (b^T Y^{(L)}(V_0))^2I\{b^T Y^{(L)}(V_0) \leq 0\}$, and by uniform integrability of the sequence $(b^T Y^{(L)}(V_n))^2I\{b^T Y^{(L)}(V_n) \leq 0\}$ this implies convergence of the first moment. Together with \eqref{eq:pa24} this establishes \eqref{eq:pa23}. Finally, the convergence
\[
\lim_{L \to \infty} \sup_{\eta \in \Gc^\eps} \Big|\|s_{\eta,\mu_0,-}^{(L)}\|_{2,\delta(\mu_0)}^2 - \|s_{\eta,\mu_0,-}\|_{2,\delta(\mu_0)}^2\Big| = 0
\]
can be proved by similar arguments as \eqref{eq:pa20} with $n^{-1}\sum_i$ replaced by the expectation, the details are omitted for the sake of brevity. This completes the proof of Assumption (A2).

\textbf{Verification of \eqref{eq:phelp1}-\eqref{eq:phelp4}} We begin by noting that for $Z \sim \eta_n$ with $\eta_n$ having support contained in $[m,M]$ it follows that $|Z-\mu_n|^k \leq M^{k-2}(Z-\mu_n)^2$ for $k \geq 3$. Thus, as soon as $\mu_n \in [m,M]$, which is the case for $n$ sufficiently large, we have
\[
\sum_{k \geq 2} a_k^2(\eta_n,\mu_n) = \sum_{k \geq 2} \frac{k^2(\E[(Z-\mu_n)^k])^2}{k!\mu_n^k } \leq (\E[(Z-\mu_n)^2])^2 \sum_{k \geq 2} \frac{ k^2 M^{2k-4}}{k!m^k } \leq C(\E[(Z-\mu_n)^2])^2.
\]
This shows \eqref{eq:phelp2}. Next, observe that
\begin{align*}
\Big(\frac{w(\eta_n,\mu_n)}{w^{(L)}(\eta_n,\mu_n)}\Big)^2 = \frac{\sum_{\ell=1}^\infty \ell^{-2}a_\ell^2(\eta_n,\mu_n)}{\sum_{\ell=1}^L \ell^{-2}a_\ell^2(\eta_n,\mu_n)} = 1 + \frac{\sum_{\ell=L+1}^\infty \ell^{-2}a_\ell^2(\eta_n,\mu_n)}{\sum_{\ell=1}^L \ell^{-2}a_\ell^2(\eta_n,\mu_n)}.
\end{align*}
Now for $Z \sim \eta_n$ with $\eta_n$ having support contained in $[m,M]$ we have as soon as $\mu_n \in [m,M]$
\begin{align*}
0 \leq \frac{\sum_{\ell=L+1}^\infty \ell^{-2}a_\ell^2(\eta_n,\mu_n)}{\sum_{\ell=1}^L \ell^{-2}a_\ell^2(\eta_n,\mu_n)}
\leq 
\frac{\sum_{k=L+1}^\infty \frac{(\E[(Z-\mu_n)^k])^2}{k!\mu_n^k }}{\frac{(\E[(Z-\mu_n)^2])^2}{2\mu_n^2 }} 
\leq
2 M^2 \sum_{k \geq L+1} \frac{ M^{2k-4}}{k!m^k } \leq CL^{-1}.
\end{align*}
The first part of \eqref{eq:phelp1} follows, and the second part of \eqref{eq:phelp1} can be established by exactly the same arguments. Finally, for $\tilde Z \sim \eta$ 
\[
\Big(\frac{w^{(L)}(\eta_n,\mu_n)}{w^{(L)}(\eta,\mu_0)}\Big)^2 = \frac{\sum_{k=1}^L \frac{(\E[(Z-\mu_n)^k])^2}{k!\mu_n^k }}{\sum_{k=1}^L \frac{(\E[(\tilde Z-\mu_0)^k])^2}{k!\mu_0^k }}
\]
and by construction $\E[(Z-\mu_n)^k] = \E[(\tilde Z-\mu_0)^k]$ for all $k \in \N$. Now \eqref{eq:phelp4} follows since $\max_{k=1,..,L} |(\mu_n/\mu_0)^k - 1| \to 0$ as $n \to \infty$. This completes all proofs for the Poisson case.  \hfill $\Box$

\end{appendix}

\end{document}